%% file: gleam_pulsars.tex
\documentclass{pasa}%

\usepackage{graphicx}
\usepackage{graphics}
\usepackage{rotating}
\usepackage{amsmath}
\usepackage{amssymb}
\usepackage{color}

\def\USydney{$^{1}$}
\def\CAASTRO{$^{2}$}
\def\UWisc{$^{3}$}
\def\CASS{$^{4}$}
\def\Curtin{$^{7}$}

\newcommand{\degree}{\ensuremath{^\circ}}
\newcommand{\thead}[1]{\multicolumn{1}{c}{#1}}
\newcommand{\changed}[1]{{#1}}

\title[Spectral energy distributions of pulsars]{Low frequency spectral energy distributions of radio pulsars detected with the Murchison Widefield Array}
\author[Murphy et al.]{

Tara Murphy\USydney$^,$\CAASTRO\thanks{E-mail: tara.murphy@sydney.edu.au},
David L. Kaplan\UWisc,
Martin E. Bell\CASS$^,$\CAASTRO,
J. R. Callingham\USydney$^,$\CASS$^,$\CAASTRO,
Steve Croft$^{5,6}$,
Simon Johnston\CASS,
Dougal Dobie\USydney,
Andrew Zic\USydney,
Jake Hughes\USydney,
Christene Lynch\USydney$^,$\CAASTRO,
Paul Hancock\Curtin$^,$\CAASTRO,
Natasha Hurley-Walker\Curtin,
Emil~Lenc$^{1,2}$,
K.~S.~Dwarakanath$^{8}$,
B.-Q.~For\Curtin,
B.~M.~Gaensler$^{1,9,2}$,
L.~Hindson$^{10}$,
M.~Johnston-Hollitt$^{10}$,
A.~D.~Kapi\'{n}ska$^{11,2}$,
B.~McKinley$^{12,2}$,
J.~Morgan\Curtin,
A.~R.~Offringa$^{13}$,
P.~Procopio$^{12,2}$,
L.~Staveley-Smith$^{11,2}$,
R.~Wayth\Curtin,
C.~Wu\Curtin,
Q.~Zheng$^{10}$
\affil{$^{1}$Sydney Institute for Astronomy, School of Physics, The University of Sydney, NSW 2006, Australia}%
\affil{$^{2}$ARC Centre of Excellence for All-sky Astrophysics (CAASTRO)}%
\affil{$^{3}$Department of Physics, University of Wisconsin--Milwaukee, Milwaukee, WI 53201, USA}%
\affil{$^{4}$CSIRO Astronomy and Space Science (CASS), Marsfield, NSW 2122, Australia}%
\affil{$^{5}$Astronomy Department, University of California, Berkeley, 501 Campbell Hall \#3411, Berkeley, CA 94720, USA}%
\affil{$^{6}$Eureka Scientific, Inc., 2452 Delmer Street Suite 100, Oakland, CA 94602, USA}%
\affil{$^{7}$International Centre for Radio Astronomy Research, Curtin University, Bentley, WA 6102, Australia}%
\affil{$^{8}$Raman Research Institute, Bangalore 560080, India}%
\affil{$^{9}$Dunlap Institute for Astronomy \& Astrophysics, University of Toronto, 50 St George St, Toronto, ON, M5S 3H4, Canada}%
\affil{$^{10}$School of Chemical \& Physical Sciences, Victoria University of Wellington, Wellington 6140, New Zealand}%
\affil{$^{11}$International Centre for Radio Astronomy Research (ICRAR), University of Western Australia, Crawley, WA 6009, Australia}%
\affil{$^{12}$School of Physics, The University of Melbourne, Parkville, VIC 3010, Australia}%
\affil{$^{13}$Netherlands Institute for Radio Astronomy (ASTRON), PO Box 2, 7990 AA Dwingeloo, The Netherlands}}%

\jid{PASA}
\doi{10.1017/pas.\the\year.xxx}
\jyear{\the\year}

\usepackage{aas_macros}
\usepackage{hyperref} 
\hypersetup{colorlinks,citecolor=blue,linkcolor=blue,urlcolor=blue}

\begin{document}

\begin{frontmatter}
\maketitle

\begin{abstract}
We present low-frequency spectral energy distributions of 60 known radio pulsars observed with the Murchison Widefield Array (MWA) telescope. We searched the GaLactic and Extragalactic All-sky MWA (GLEAM) survey images for 200-MHz continuum radio emission at the position of all pulsars in the ATNF pulsar catalogue. For the 60 confirmed detections we have measured flux densities in $20\times8$~MHz bands between 72 and 231 MHz. We compare our results to existing measurements and show that the MWA flux densities are in good agreement. %These observations will be helpful in planning future pulsar timing observations with low-frequency radio telescopes.
\end{abstract}

\end{frontmatter}

\section{INTRODUCTION}
\label{sec:intro}
Pulsars are generally observed at high time resolution in order to detect and resolve their pulses. However, many can also be detected in continuum interferometric images via their phase-averaged emission \citep[e.g.,][]{kaplan98} which offers additional useful information. For example, in contrast to difficulties in absolute flux calibration of single dish observations \citep[e.g.,][]{lorimer12}, interferometers allow accurate flux density measurements that can be used to help constrain  pulsar emission mechanisms \citep{malofeev80,lorimer95,karastergiou15} and derive the pulsar luminosity function, and hence Galactic pulsar birth rate \citep{lorimer93}. Separately, interferometry can be used to determine accurate positions (and eventually proper motions and parallaxes) for pulsars \citep[e.g.,][]{chatterjee09,deller09,deller11,deller16} to aid or augment pulsar timing \citep[e.g.,][]{gaensler99,lorimer12}.

Pulsars have been observed across the entire electromagnetic spectrum, from frequencies as low as 10~MHz \citep{hassall12} up to 1.5 TeV \citep{magic16}. At radio frequencies the spectral behaviour of the majority of pulsars can be described by a power law of the form $S_\nu \propto \nu^\alpha$ where $\alpha$ is the spectral index and $S_\nu$ is the flux density at frequency $\nu$ \citep{lorimer95}. Pulsars typically have steep spectra with the mean spectral index of non-recycled pulsars around $\alpha = -1.6$ \citep[e.g.][]{sieber73,lorimer95} and millisecond pulsars (MSPs) around $\alpha = -1.8$ \citep[e.g.][]{kramer99,maron00}. Simulations by \citet{bates13} that take into account selection biases present in the distributions of known pulsars show the underlying spectral index is likely to be $\alpha \approx -1.4$.

\changed{Typical single-dish pulsar observations can have poor flux calibration (as bad as $\sim 50$\%) due to an absence of reliable calibration sources, unknown positions in primary beams (at least initially)}, and the difficulties in calibrating single-dish telescopes (see \citealt{oneil02,lorimer12}). Even observations with arrays that have been coherently beamformed can have poor flux calibration, since calibration relies on modelling and characterising the performance across the fields-of-view and over time, often without the ability to verify and track the performance during the science observation. \changed{These issues are discussed by, for example, \citet{bilous16}.} In contrast, imaging observations can be less sensitive \changed{(limited both by confusion, and also by
the duty cycle of the pulsations)} but can be calibrated very robustly through simultaneous observations of hundreds or thousands of sources. 

\changed{Only a small fraction of radio detected pulsars have had their continuum flux densities measured across a range of frequencies; for example in the ATNF Pulsar Catalogue 1919 out of 2573 sources have zero or one radio continuum flux density measurements listed.}
Low frequency observations of pulsars are important for investigating their
spectral indices, and in particular in determining the frequency of their spectral turnovers. Spectral turnovers have been observed in a small number of sources at frequencies below 400~MHz \citep[e.g.][]{sieber73,ellingson13} and in roughly $10\%$ of pulsars in total \citep{maron00}. The cause of spectral turnovers is not known, but it is thought to be either
synchrotron self-absorption in the emission region, or thermal absorption by a gas cloud in the line of sight to a pulsar \citep{sieber73}. One of the aims of a low frequency study of pulsars is to collect a larger sample of sources that exhibit spectral turnovers and hence help address this question by relating the presence of turnovers to other intrinsic or extrinsic parameters.

With good spectral coverage, continuum observations of pulsars can be used for modelling pulsar emission mechanisms, and studying the statistical properties of pulsar populations in a way that is independent of their time-varying properties such as period and dispersion measure \citep{lorimer95}. Continuum observations have the advantage that they are less susceptible to interstellar propagation effects (dispersion and scattering) that smear out the emission over the pulse
phase.   The reliability of  flux density measurements can be adversely affected by interstellar scintillation, which is very strong at low frequencies \citep{rickett77,bhat14}.  However, for most pulsars the scintillation bandwidth will be much smaller than our observation bandwidth so that the scintillation will saturate: only for the nearest pulsars will a finite number of scintles cause significant variability \citep{bell16}.

%The timescale for refractive scintillation is inversely proportional to the observing frequency \citep{rickett77}, and the timescale for scattering scales as inverse frequency to the fourth power, so these effects are particularly important at low frequencies \citep[e.g.][]{bhat14}.

Low frequency continuum studies of pulsars have been conducted by a number of groups. \citet{kouwenhoven00} measured Westerbork Northern Sky Survey (WENSS) 325~MHz flux densities for 39 pulsars and \citet{kuniyoshi15} provide 74~MHz VLA Low Frequency Sky Survey, redux \citep[VLSSr;][]{lane14} and 325~MHz WENSS flux densities for 10 millisecond pulsars detected in the VLSSr.
\citet{kaplan00} reported 365~MHz flux densities for 6 pulsars in the Texas survey catalogues \citep{douglas96}.
Two recent LOw-Frequency ARray \citep[LOFAR;][]{vanhaarlem13} projects have measured low frequency fluxes for 158 non-recycled (or normal) pulsars \citep{bilous16} and 48 millisecond pulsars \citep{kondratiev16}. Most recently \citet{frail16} made 150~MHz measurements of 200 known pulsars using data from the reprocessed GMRT Sky Survey \citep[TGSS ADR;][]{intema16}. In this paper we present Murchison Widefield Array observations of 60 pulsars at frequencies between 72 and 231~MHz.
 
\section{OBSERVATIONS AND DATA REDUCTION}\label{s_obs}
\subsection{Observations}
The Murchison Widefield Array \citep[MWA;][]{tingay13} is a 128-tile low-frequency radio interferometer located in Western Australia. One of the major MWA projects is the GaLactic and Extragalactic All-sky MWA survey \citep[GLEAM;][]{wayth15}. GLEAM is a survey of the radio sky south of declination $+30^\circ$ at frequencies between 72 and 231~MHz, conducted between 2013 June and 2014 July. The survey was performed using five instantaneous observing bandwidths of 30.72\,MHz, with each band observing the same part of the sky for an integration time of $\sim$2 minutes. The observing bands were further subdivided into four sub-bands with bandwidths of 7.68\,MHz during processing. Hence, the GLEAM survey reports twenty flux density measurements between 72 and 231\,MHz. 

Note that although the entire sky was imaged in the GLEAM survey, the first major GLEAM catalogue release \citep{hurleywalker16} excludes the Galactic plane region ($|b|<10^\circ$). However, in this work we measure flux densities directly from the survey images and hence cover the whole $\delta<30^\circ$ sky.

\subsection{Data reduction}
The data reduction process that was performed is discussed in detail by \citet{hurleywalker16}. In summary, the raw visibility data from the MWA observations were processed by \textsc{Cotter} \citep{offringa15} and radio frequency interference (RFI) was excised using the \textsc{AOFlagger} algorithm \citep{offringa12}. For the five instantaneous observing
bandwidths of 30.72\,MHz, an initial model of the sky was used to apply initial amplitude and phase calibration solutions. Imaging was performed using \textsc{WSClean} \citep{offringa14}, with a ``robust'' parameter of $-1.0$ (close to uniform weighting). \changed{Uniform weighting weights the visibilities in inverse proportion to the sampling density function. This has the effect of minimising the sidelobe level and hence minimises contamination from diffuse structure and to aid in easily identifying unresolved sources, such as pulsars.} Multi-frequency synthesis was applied across the instantaneous
bandwidth for each snapshot observation, and \textsc{clean}ed \citep{hogbom74} to the first negative \textsc{clean} component. The observations were then divided into four 7.68-MHz sub-bands and jointly \textsc{clean}ed, resulting in a RMS of $\sim$250 to $\sim$50\,mJy~beam$^{-1}$ for 72 to 231\,MHz, respectively. The 7.68~MHz sub-band images were then put through a self-calibration loop, using the initial calibrator images to ensure position and flux density consistency and stability.

An initial flux density scale for the images was then set using the Molonglo Reference Catalogue \citep[MRC;][]{large81,large91}, scaled to the respective frequency, and an astrometric correction was applied using the sources referenced in MRC. The snapshots for an observed declination strip were mosaicked, with each snapshot weighted by the square of the primary beam response. Any residual declination dependence of the flux density scale in the mosaics, due to uncertainties in the primary beam model, was corrected using the VLSSr, MRC, and NRAO VLA Sky Survey \citep[NVSS;][]{condon98} catalogues. We estimate that the flux density calibration is accurate to $8\%$ for sources with $|b| > 10\degree$ and up to $20\%$ for sources with $|b| < 10\degree$.

A deep wide-band image covering $170-231$~MHz was formed for each mosaic. The deep wide-band image provides a higher signal-to-noise ratio and more accurate source positions than what can be attained for a single 7.68-MHz sub-band image. The \textsc{Background And Noise Estimator (bane)}\footnote{\texttt{https://github.com/PaulHancock/Aegean/wiki/BANE}} was used to measure the background and noise properties of the deep-wideband images. \textsc{bane} estimates the background and noise of an image as the median and standard deviation of the pixel distribution over a sliding window. 
\changed{Calculating the background and noise properties in this way is biased by the presence of sources. \textsc{bane} mitigates this bias via sigma clipping of the pixel distribution (3 rounds of 3$\sigma$).}
\textsc{bane} creates maps of the sky of the same dimensions as the input image, with each pixel representing either the background or noise level at a given location. The background and noise maps were then passed to the source finding and characterisation program \textsc{Aegean} v1.9.6 \citep{hancock12} to form a reference catalogue. The positions of the sources in the reference catalogue were then convolved with the appropriate synthesised beam at each sub-band frequency to characterise the flux density of the sources in each of the 20 sub-band images.

\subsection{Sample Selection}
\begin{figure*}
\includegraphics[width=\columnwidth]{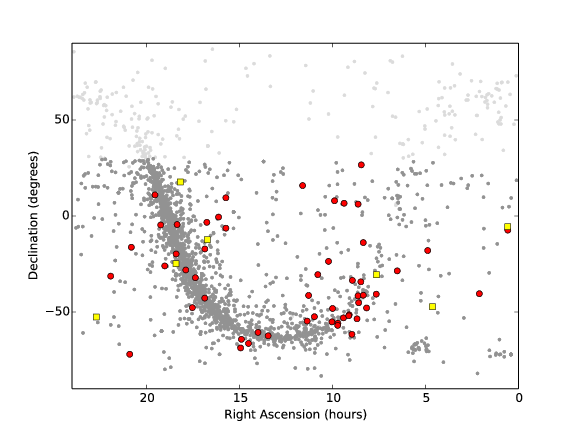}
\includegraphics[width=\columnwidth]{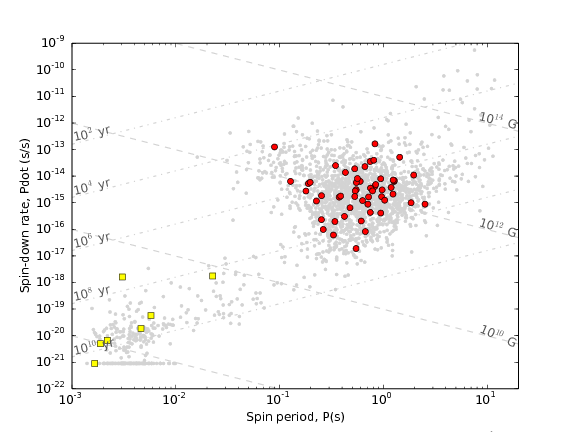}
\caption{{\it Left:} Distribution of all known pulsars from the ATNF pulsar catalogue (light grey dots); the pulsars observable by GLEAM (dark grey dots) and the pulsars detected in GLEAM and presented in this paper: millisecond pulsars are shown as yellow squares, and non-recycled pulsars as red circles. The Galactic plane, where most known pulsars lie, is clearly visible.
{\it Right:} Distribution of all known pulsars (pale grey dots) and detected pulsars on the $P-\dot{P}$ diagram. Pulsars with an unknown $\dot P$ but with $P<0.01$~s are plotted at $\dot P = 10^{-21}$ s s$^{-1}$ (this includes PSR~J1810$+$1744).  We also show contours of constant dipole magnetic field and spin-down age, as labelled.}
\label{distrib}
\end{figure*}

We selected all sources from the ATNF pulsar catalogue v1.54 \citep{manchester05} that fell within the observed GLEAM region of $\delta < +30^\circ$. We excluded globular cluster pulsars, and those with a positional uncertainty of greater than 1 arcmin. This left a sample of 1996 sources. We then searched the GLEAM 170--231~MHz mosaics for $3\sigma$ detections within 2 arcmins of the positions of these sources. We manually inspected postage stamp images of the potential detections, ruling out artefacts and coincident
extragalactic sources based on their visual morphology. We also ruled out sources with existing non-pulsar identifications in SIMBAD or the NASA/IPAC Extragalactic Database.

This resulted in a sample of 60 sources with GLEAM detections, as shown in Figure~\ref{distrib}. Using a definition of millisecond pulsars as having spin periods of $P<30$~ms and spin-down rates of $\dot P <1\times10^{-16}$~s~s$^{-1}$, six sources in our sample are millisecond pulsars and 53 are normal pulsars. One source, PSR~J1810$+$1744, has an unknown spin-down rate, but its spin period of 0.00166~s indicates it is a millisecond pulsar. For each of these candidate sources we measured the flux density at the position of the pulsar using the source finding package {\sc Aegean} v1.9.6 \citep{hancock12}.
\changed{The measured MWA positions agreed well with the positions in the ATNF pulsar catalogue; the mean position offsets are $\Delta \alpha = 13^{\prime\prime}$, $\Delta \delta = 9^{\prime\prime}$.}

\subsection{False detection rate}
It is possible that some of the matches with ATNF pulsar positions are due to chance coincidences. To estimate what fraction this might be, we can consider the areal density of sources in the GLEAM catalogue.
The GLEAM catalogue has 304\,894 sources over an area of 24\,402 square degrees, or $\sim 12$ sources per square degree. Therefore any given pulsar has a $3\%$ chance of being matched with a background source by chance coincidence. This implies that out of the detections we made, it is possible that $1-2$ of them are false positives.

The actual coincident source rate could be somewhat higher than this, since we searched to a lower flux density cutoff than the published GLEAM catalogue, and we also searched in the Galactic plane where the source density is higher. However, we expect the false detection rate to be of order a few sources.

\section{Results and Discussion}\label{s_results}
\begin{figure}[t!]
\includegraphics[width=\columnwidth]{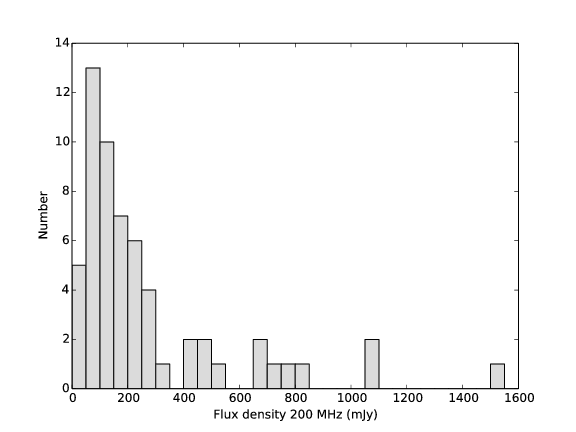}
\caption{Distribution of GLEAM 200-MHz flux density measurements for the 60 sources in our sample.}
\label{fhist}
\end{figure}

We fit and parameterised each of the 60 detected sources in the GLEAM averaged mosaics (centred on 200~MHz) using {\sc Aegean}'s priorised\footnote{Priorised fitting holds the position and shape of a source fixed and only fits for the flux density, thus making it possible to measure the flux densities of sources that are below the nominal detection limit and using prior information to reduce the uncertainties in subsequent measurements.} fitting option. The distribution of measured flux densities is shown in Figure~\ref{fhist}. We then fit a source at the same position in each of the $20\times 8$~MHz sub-band images. We measured flux densities in each sub-band image if there was a detection above $3\sigma$. We excluded sources that were obviously extended (by visual inspection), and so were able to fit each source with a point source model.

\subsection{Comparison with the literature}
Fifteen of the sources in our sample have a detection in either the
VLSS \citep{cohen07,lane14}, LOFAR \citep{bilous16}, \citet{slee95} or \citet{malofeev00} surveys.

The \citet{bilous16} LOFAR census provides a good comparison as the mean frequency of the LOFAR High-Band Antennas (HBA)
is 149~MHz which aligns well with our GLEAM sub-band centred on 151~MHz.
The \citet{bilous16} flux densities were measured from the single best observation (each observation covered at least 1000 spin periods), and their comparisons with LOFAR imaging observations suggest there may be up to $40\%$ difference between the flux densities obtained through these different methods. The agreement to our data is roughly within these limits.

Four of our sources are in the \citet{bilous16} sample:
\begin{itemize}
\item {\bf J0826$+$2637 (B0823$+$26)} has an MWA 151~MHz flux density of $365\pm48$~mJy. The LOFAR 149~MHz flux density is somewhat higher at $522\pm261$mJy, but still agrees within the stated errors of both surveys;
\item {\bf J1136$+$1551 (B1133$+$16)} has an MWA 151~MHz flux density of $1057\pm49$~mJy, which agrees well with the LOFAR 149~MHz flux density of $935\pm467$~mJy;
\item {\bf J1543$+$0929 (B1541$+$09)} has an MWA 151~MHz flux density of $371\pm45$~mJy, which is substantially lower than the LOFAR 149~MHz flux density of $768\pm384$~mJy, although still within the reported uncertainty of
the LOFAR measurement.
\end{itemize}
We do not have a reliable sub-band flux density measurement for {\bf J1932$+$1059 (B1929$+$10)} but our 200~MHz
averaged band measurement agrees reasonably well with existing literature measurements, including LOFAR (see the plot in Figure~\ref{seds1}). Inspection of the spectral energy distribution suggests our MWA results are overall in good agreement with the rest of the literature measurements, for example the TGSS~ADR1 150~MHz flux density is $368\pm41$~mJy.

In addition, three of our millisecond pulsars are in the \citet{kondratiev16} sample, which presents flux densities, also averaged over the 110--188~MHz band (centred on 149~MHz). Note that the errors quoted in Table~4 of \citet{kondratiev16} are much smaller than the actual estimated errors of $50\%$, so for this comparison we have assumed $50\%$ errors on the LOFAR flux densities:
\begin{itemize}
\item {\bf J0034$-$0534} has an MWA 151~MHz flux density of $394\pm24$~mJy which agrees with the LOFAR 149~MHz flux density of $491\pm245$~mJy;
\item {\bf J0737$-$3039A} has an MWA 200~MHz flux density of $53\pm8$~mJy and an MWA 143~MHz flux density of $89\pm25$~mJy, both of which agree with the LOFAR 149~MHz flux density of $64\pm32$~mJy;
\item {\bf J1810$+$1744} has an MWA 151~MHz flux density of $320\pm102$~mJy which agrees within the errors with the LOFAR 149~MHz flux density of $563\pm282$~mJy.
\end{itemize}

\subsection{Spectral energy distributions}
\begin{figure*}
\includegraphics[width=0.65\columnwidth]{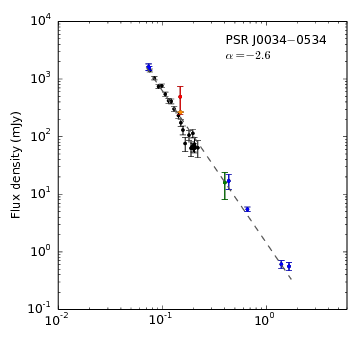}\hspace*{-0.5em}%                     
\hspace*{-0.5em}\includegraphics[width=0.65\columnwidth]{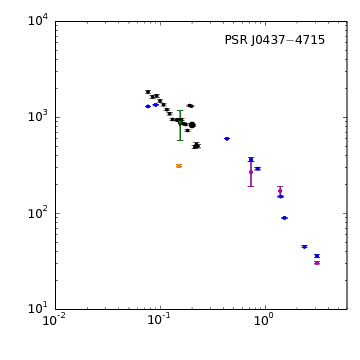}\hspace*{-0.5em}%     
\hspace*{-0.5em}\includegraphics[width=0.65\columnwidth]{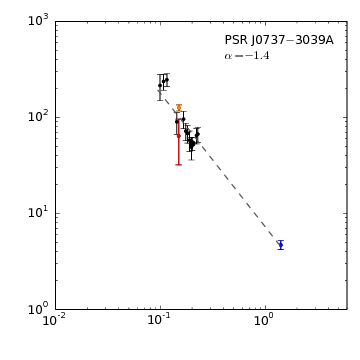} \\[-0.6em]
\includegraphics[width=0.65\columnwidth]{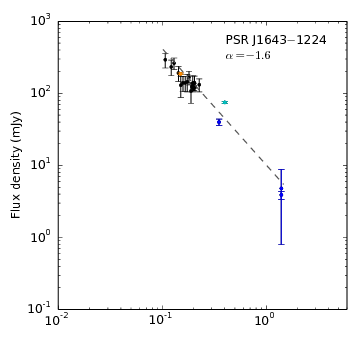}\hspace*{-0.5em}%                     
\hspace*{-0.5em}\includegraphics[width=0.65\columnwidth]{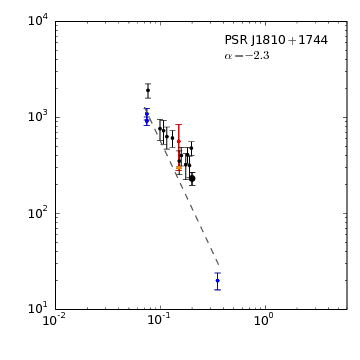}\hspace*{-0.5em}%     
\hspace*{-0.5em}\includegraphics[width=0.65\columnwidth]{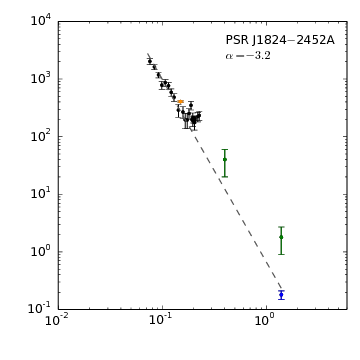} \\[-0.6em]
\includegraphics[width=0.65\columnwidth]{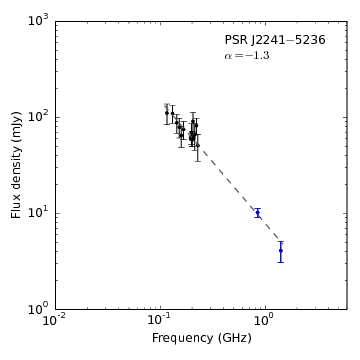}\hspace*{-0.5em}%                     
\raisebox{0.9cm}{\hspace*{1.25cm}\includegraphics[width=0.4\columnwidth]{images/legend}}\\
\caption{Spectral energy distributions for millisecond pulsars in our sample. PSR~J0437$-$4715 could not be fit by a single or broken power law, and is known to be highly variable due to scintillation.}
\label{seds2}
\end{figure*}

In Figures~\ref{seds1} and \ref{seds2} we present spectral energy distributions (SEDs) for each of the normal and millisecond pulsars in our sample, respectively. Where it was possible to measure sub-band flux densities with greater than $3\sigma$ significance we have included the individual sub-band flux densities in the SEDs (for example, PSR~J0034$-$0721). In other cases the sub-band measurements were either below the noise, or we excluded them due to calibration issues with Galactic plane data (for example, PSR~J0630$-$2834). In these cases we only included the 200~MHz flux density measured from the 170--231~MHz mosaics.

It is important to note that the literature fluxes come from many different projects, with different observational setups. In addition, low frequency flux density measurements are more affected by scintillation than higher frequency measurements, and pulsars can also be intrinsically variable. As a result, fluxes measured by different groups at different times
may vary by an order of magnitude. We have excluded two sets of recent measurements \citep{stovall15,frail16} from our fits (but included them in the SEDs) as the flux calibration requires further investigation\footnote{See discussion of flux density scale at: http://tgssadr.strw.leidenuniv.nl}.

We fit each of our pulsar spectral energy distributions with both a
single power law of the form:
\begin{equation}
S_{\nu} = S_{\nu_0} \left( \frac{\nu}{\nu_0} \right)^\alpha
\end{equation}
where $\alpha$ is the spectral index, and $S_{\nu_0}$ is the flux density at a reference frequency $\nu_0$; and a broken power law of the
form:
\begin{equation}
S_{\nu} =
\begin{cases}
S_{\nu_0} \left( \frac{\nu}{\nu_0} \right)^{\alpha_{\rm lo}} &\text{if}\; \nu < \nu_{\rm br} \\
S_{\nu_0} \left( \frac{\nu_{\rm br}}{\nu_0} \right)^{\alpha_{\rm lo}} \left( \frac{\nu}{\nu_{\rm br}\
} \right)^{\alpha_{\rm hi}} &\text{if}\; \nu > \nu_{\rm br} \\
\end{cases}
\end{equation}
where $\nu_{\rm br}$ is the break frequency, and $\alpha_{\rm lo}$ and $\alpha_{\rm hi}$ are the spectral indices below
and above that break frequency, respectively.

Model fits to the SEDs were conducted using a nonlinear least-squares routine that applied the Levenberg-Marquardt algorithm, an iterative procedure that linearizes the function at each step based on a new estimate of the function from the gradient of the previous step. The fitting routine produced a covariance matrix with which 1-$\sigma$ uncertainties on the parameters were taken to be the square-root of the diagonal terms. In the fitting procedure the uncertainties on the data points were assumed to be independent and Gaussian. Initial conditions for the broken power-law fits were selected by random sampling using a Monte Carlo simulation.
In cases where there were not enough data points to support a broken power-law fit we chose a single power law. In cases where the reduced $\chi^2$ suggested a broken power-law fit was preferred we ran an F-test and rejected the null hypothesis for probability $P < 0.01$.

\begin{figure}
\includegraphics[width=\columnwidth]{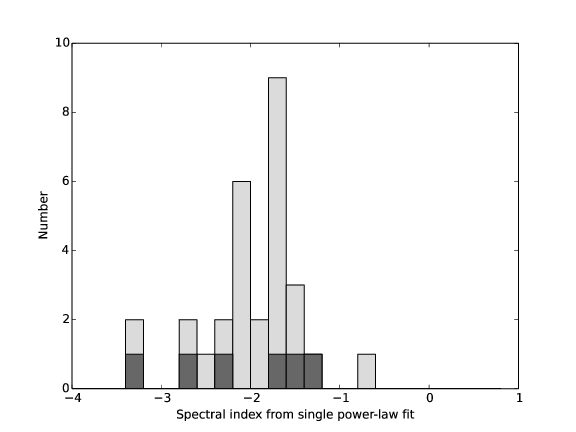}
\caption{Distribution of spectral indices for sources for which the SEDs were fit by a single power law. The dark grey shading shows the millisecond pulsars.}
\label{f_sihist}
\end{figure}

We found that 28 of the pulsars in our sample could be fit by a single power law. The distribution of spectral indices from these fits is shown in Figure~\ref{f_sihist}. The individual fits are shown by the dashed lines in Figures~\ref{seds1} and \ref{seds2} and the spectral indices derived from these fits are
listed in Tables~\ref{sinormal} and \ref{similli}.
30 of the remaining sources were fit by a broken power law and the results from these fits are given in Table~\ref{broken}. We note that some of these sources show signs of a spectral turnover whereas others have a clear spectral break. The flux density measurements for each GLEAM sub-band are
given in Table~4.
%\ref{t_seds}

One source (PSR~J0828$-$3417) has too few points to be fit. Another source (PSR~J0437$-$4715) could not be fit by either a single or broken power law. This source is known to scintillate \citep{bell16} and is discussed further in Section~\ref{s_var}. \changed{The GLEAM sub-band points show this clearly, with significant changes in the flux density between neighbouring bands. The GLEAM observations cycle through each of the five major frequency bands, and so the observations at different frequencies are not simultaneous.} This variability means the SED can not be well fit.

PSR~J0942$-$5657 is not fit well due to several points with higher than
expected flux densities at low frequencies. This could be due to the low
frequency measurement, which is at relatively low resolution, picking up
diffuse emission from a surrounding supernovae remnant or pulsar wind
nebula. It is not possible to disentangle these factors with the MWA continuum data alone.

Currently our sample is too small to see if the spectral indices we measure here correlate with any intrinsic parameters of the pulsars (such as spin-period, spin-down age, dipole magnetic field, energy-loss rate), but we expect that with deeper surveys in the near future we can increase the number of pulsars significantly (see Section~\ref{s_conclusion}) and will be more sensitive to population-wide trends.

\begin{figure}
\includegraphics[width=\columnwidth]{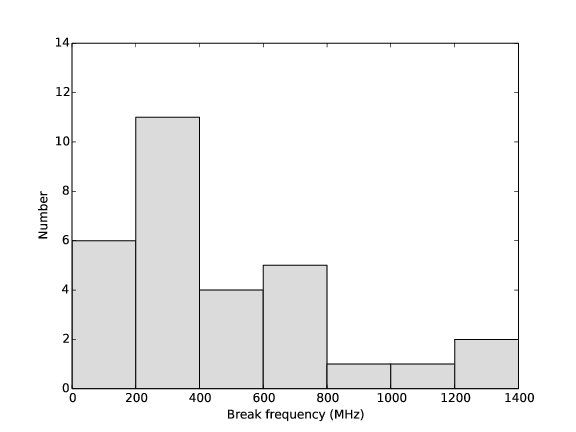}
\caption{The distribution of spectral break frequencies for pulsars that were fit by a broken power law.}
\label{f_break}
\end{figure}

Previous work has shown that approximately $10\%$ of pulsars can not be fit by a single power law \citep{maron00}.
We found a substantially higher fraction than this ($30/58 = 52\%$).
The higher percentage is probably due to our sample being able to detect sources with low frequency spectral breaks. The distribution of spectral break frequencies for pulsars that were fit by a broken power law is shown in Figure~\ref{f_break}: 23 of our sources have $v_{\rm br} \lesssim 400$~MHz. The sources with spectral breaks do not have any obvious intrinsic parameter that would select for this property.

\begin{table*}
  \centering
  \caption{Flux density measurements and spectral indices for the non-recycled pulsars in our sample\
. The flux density at 200~MHz (S$_{200}$) is measured from mosaics that are averaged across the full 72--231~MHz bandwidth.  Sources
identified as variable by \citet{bell16} are marked with (v). See Table~\ref{broken} for sources with
broken power law fits (listed as `2pl' in this table). $N_{\rm p}$ is the number of measurements included in the fit. Full SEDs are given in Table~4. Note that \citet{stovall15} and \citet{frail16} measurements were not included in our fits, as discussed in the text.}
  \label{sinormal}
  \begin{tabular}{llrccrr}                 
    \hline
    J name & B name & S$_{200}$ & Fit range & $N_{\rm p}$ & $\alpha_{\rm fit}$ & SED references \\
           &        & (mJy)       & (MHz)       & &     &   \\
    \hline
\input{pulsar_1pl}

    \hline
  \end{tabular}
References:
B95 \citep{becker95}, Be16 \citep{bell16}, B16 \citep{bilous16},
Bh16 \citep{bhattacharyya16},
C07 \citep{cohen07}, C98 \citep{condon98}, D96 \citep{douglas96},
D02 \citep{debreuck02}, D15 \citep{dai15}, F16 \citep{frail16},
G93 \citep{griffith93}, G98 \citep{gould98}, H92 \citep{helfand92},
H11 \citep{hessels11}, H04 \citep{hobbs04}, H14 \citep{hurleywalker14}, I16 \citep{intema16}, J93 \citep{johnston93}, J05 \citep{johnston05}, J15 (Johnston, pc), J17 (Jankowski et al. 2017, in prep), K11 \citep{keith11}, K15 \citep{kuniyoshi15}, K16 \citep{kondratiev16}, L14 \citep{lane14},
L95 \citep{lorimer95}, L98 \citep{lyne98}, M78 \citep{manchester78}, M96 \citep{mcconnell96}, M00 \citep{malofeev00}, M03 \citep{mauch03}, M07 \citep{murphy07}, M13 \citep{manchester13},
N04 \citep{nord04}, N08 \citep{noutsos08}, R97 \citep{ramachandran97},
R10 \citep{renaud10}, S95 \citep{slee95}, Se95 \citep{seiradakis95}, T93 \citep{taylor93},
T98 \citep{toscano98}, Z13 \citep{zakharenko13}.
\end{table*}

\begin{table*}\ContinuedFloat
  \centering
  \caption{(continued)}
    \begin{tabular}{llrccrr} % four columns, alignment for each                                        
    \hline
    J name & B name & S$_{200}$ & Fit range & $N_{\rm p}$ & $\alpha_{\rm fit}$ & SED references \\
           &        & (mJy)       & (MHz)       & &     &   \\
    \hline
\input{pulsar_1plb}
    \hline
  \end{tabular}
\end{table*}

\begin{figure*}
\includegraphics[width=0.65\columnwidth]{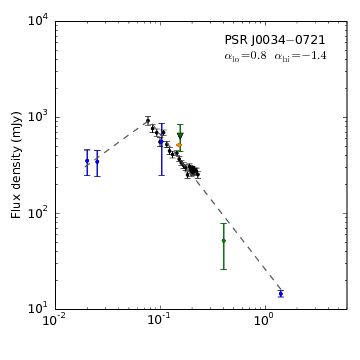}\hspace*{-0.5em}%                     
\hspace*{-0.5em}\includegraphics[width=0.65\columnwidth]{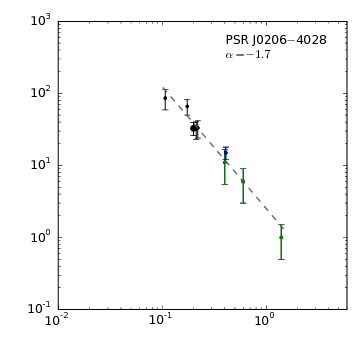}\hspace*{-0.5em}%     
\hspace*{-0.5em}\includegraphics[width=0.65\columnwidth]{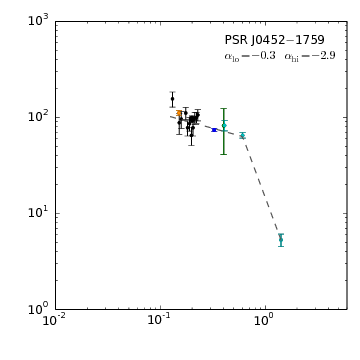}\\[-0.6em]
\includegraphics[width=0.65\columnwidth]{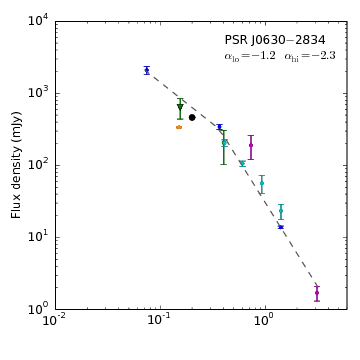}\hspace*{-0.5em}%                     
\hspace*{-0.5em}\includegraphics[width=0.65\columnwidth]{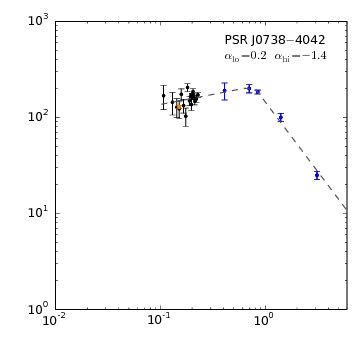}\hspace*{-0.5em}%     
\hspace*{-0.5em}\includegraphics[width=0.65\columnwidth]{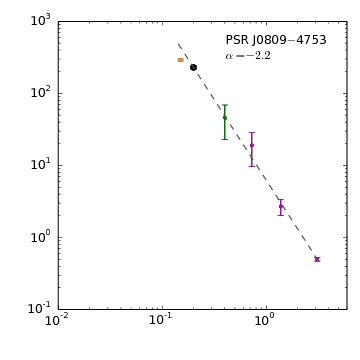}\\[-0.6em]
\includegraphics[width=0.65\columnwidth]{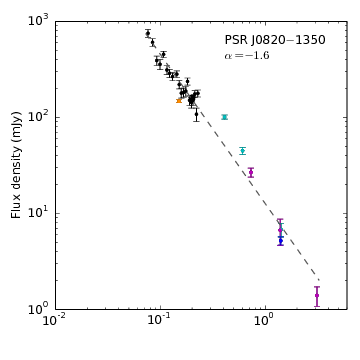}\hspace*{-0.5em}%                     
\hspace*{-0.5em}\includegraphics[width=0.65\columnwidth]{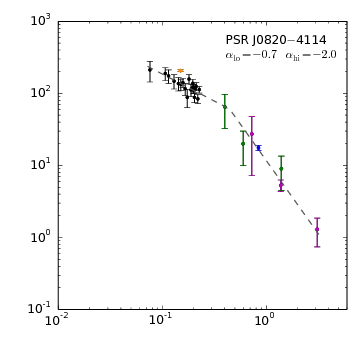}\hspace*{-0.5em}%     
\hspace*{-0.5em}\includegraphics[width=0.65\columnwidth]{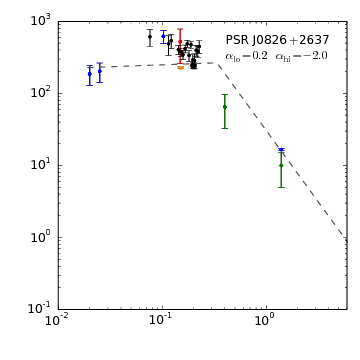}\\[-0.6em]
\includegraphics[width=0.65\columnwidth]{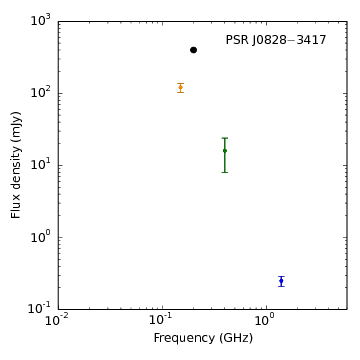}\hspace*{-0.5em}%                     
\hspace*{-0.5em}\includegraphics[width=0.65\columnwidth]{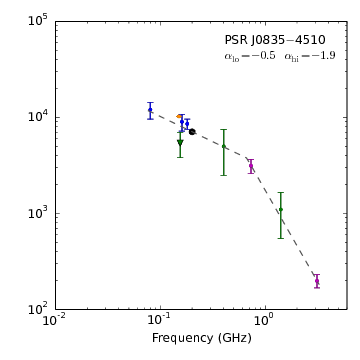}\hspace*{-0.5em}%     
\raisebox{0.9cm}{\hspace*{1.25cm}\includegraphics[width=0.4\columnwidth]{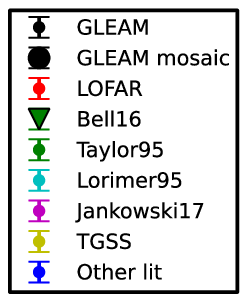}}\\
\caption{Spectral energy distributions for non-recycled pulsars in our sample. New measurements from this work are in black.
Flux density measurements from the literature are coloured according to the caption, with specific references given in
Table~\ref{sinormal}. The dashed lines show the best-fit power law (or broken power law) as discussed in the text. PSR~J0828$-$3417 had too few points to be fit.}
\label{seds1}
\end{figure*}

\begin{figure*}\ContinuedFloat
\includegraphics[width=0.65\columnwidth]{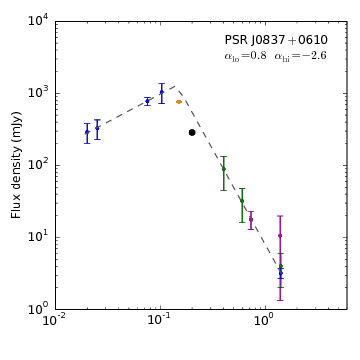}\hspace*{-0.5em}%                     
\hspace*{-0.5em}\includegraphics[width=0.65\columnwidth]{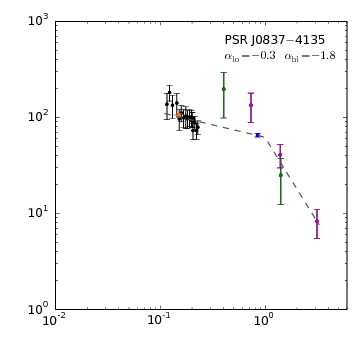}\hspace*{-0.5em}%     
\hspace*{-0.5em}\includegraphics[width=0.65\columnwidth]{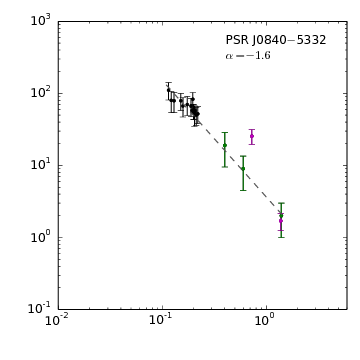} \\[-0.6em]
\includegraphics[width=0.65\columnwidth]{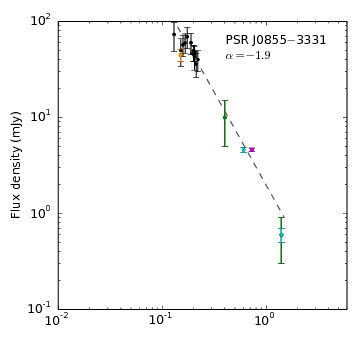}\hspace*{-0.5em}%                     
\hspace*{-0.5em}\includegraphics[width=0.65\columnwidth]{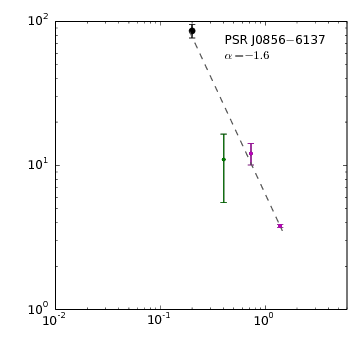}\hspace*{-0.5em}%     
\hspace*{-0.5em}\includegraphics[width=0.65\columnwidth]{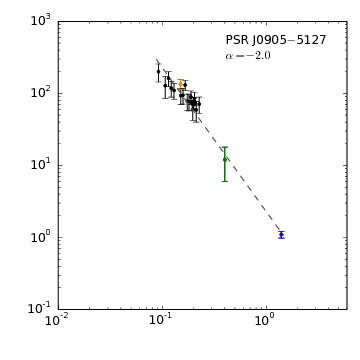} \\[-0.6em]
\includegraphics[width=0.65\columnwidth]{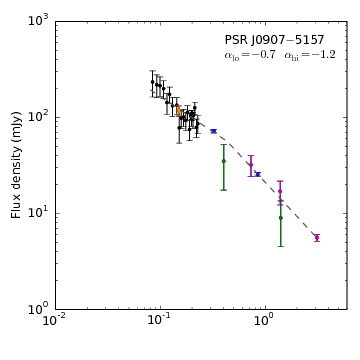}\hspace*{-0.5em}%                     
\hspace*{-0.5em}\includegraphics[width=0.65\columnwidth]{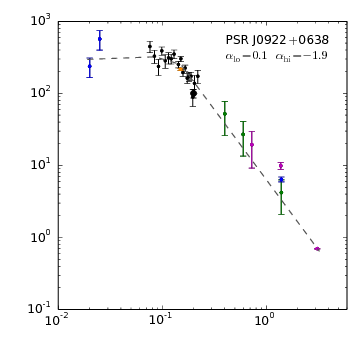}\hspace*{-0.5em}%     
\hspace*{-0.5em}\includegraphics[width=0.65\columnwidth]{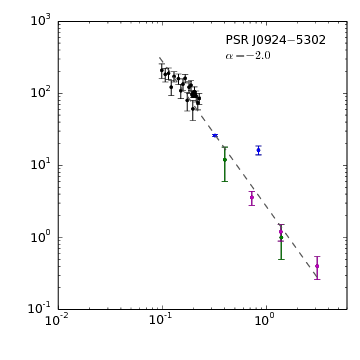} \\[-0.6em]
\includegraphics[width=0.65\columnwidth]{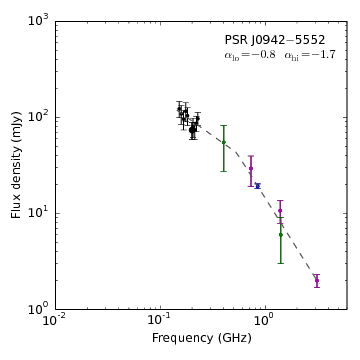}\hspace*{-0.5em}%                     
\hspace*{-0.5em}\includegraphics[width=0.65\columnwidth]{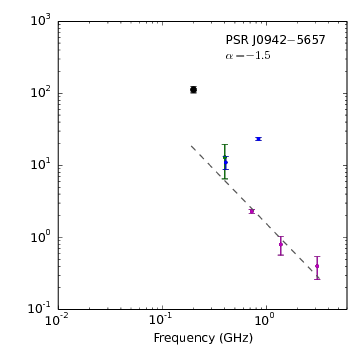}\hspace*{-0.5em}%     
\raisebox{0.9cm}{\hspace*{1.25cm}\includegraphics[width=0.4\columnwidth]{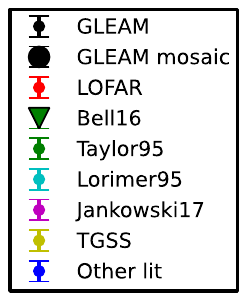}}\\
\caption{(continued)}
\end{figure*}

\begin{figure*}\ContinuedFloat
\includegraphics[width=0.65\columnwidth]{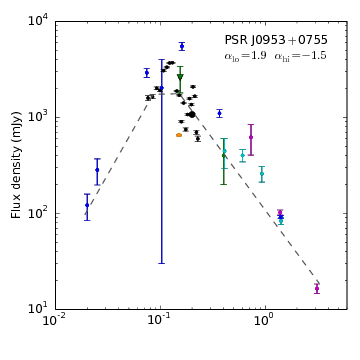}\hspace*{-0.5em}%                     
\hspace*{-0.5em}\includegraphics[width=0.65\columnwidth]{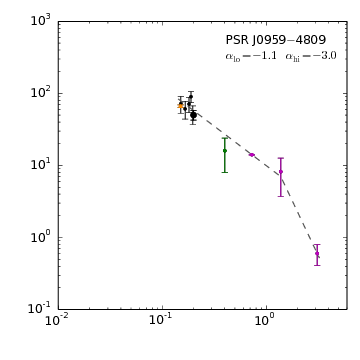} \hspace*{-0.5em}%    
\hspace*{-0.5em}\includegraphics[width=0.65\columnwidth]{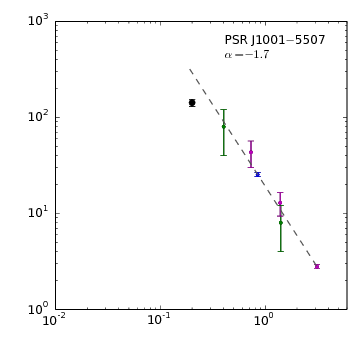} \\[-0.6em]
\includegraphics[width=0.65\columnwidth]{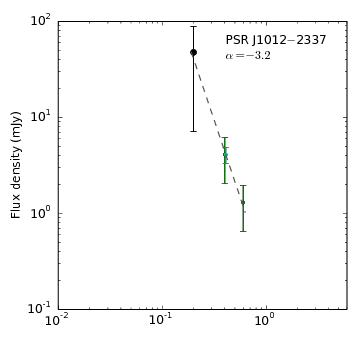}\hspace*{-0.5em}%                     
\hspace*{-0.5em}\includegraphics[width=0.65\columnwidth]{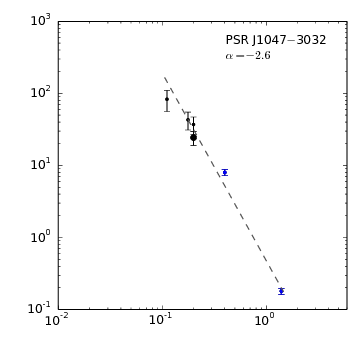}\hspace*{-0.5em}%     
\hspace*{-0.5em}\includegraphics[width=0.65\columnwidth]{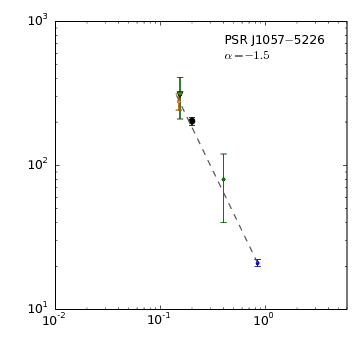} \\[-0.6em]
\includegraphics[width=0.65\columnwidth]{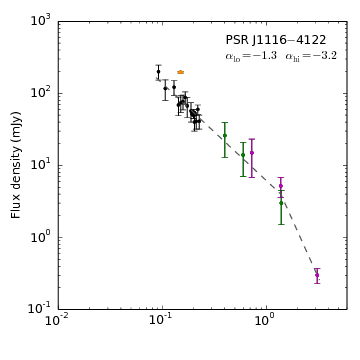}\hspace*{-0.5em}%                     
\hspace*{-0.5em}\includegraphics[width=0.65\columnwidth]{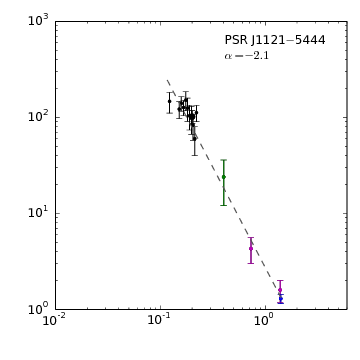}\hspace*{-0.5em}%     
\hspace*{-0.5em}\includegraphics[width=0.65\columnwidth]{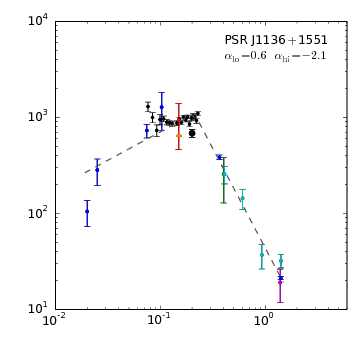} \\[-0.6em]
\includegraphics[width=0.65\columnwidth]{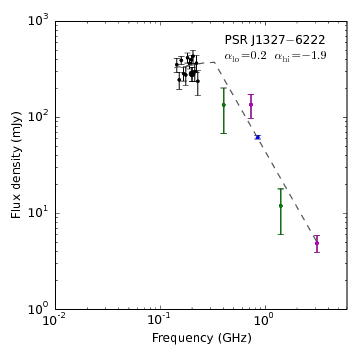}\hspace*{-0.5em} %                    
\hspace*{-0.5em}\includegraphics[width=0.65\columnwidth]{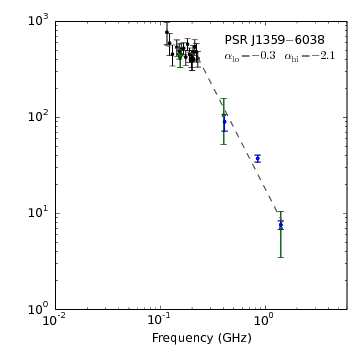}\hspace*{-0.5em}%
\raisebox{0.9cm}{\hspace*{1.25cm}\includegraphics[width=0.4\columnwidth]{images/legend}}\\
\caption{(continued)}
\end{figure*}

\begin{figure*}\ContinuedFloat
\includegraphics[width=0.65\columnwidth]{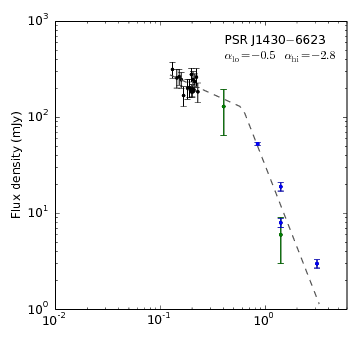}\hspace*{-0.5em}%                     
\hspace*{-0.5em}\includegraphics[width=0.65\columnwidth]{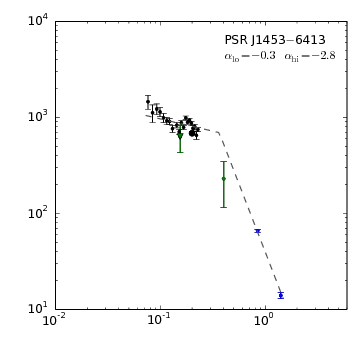}\hspace*{-0.5em}%     
\hspace*{-0.5em}\includegraphics[width=0.65\columnwidth]{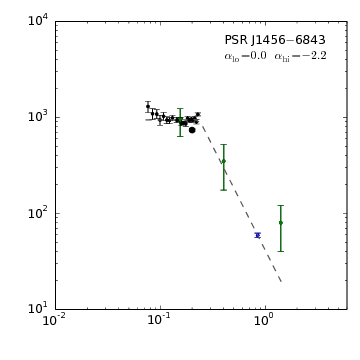} \\[-0.6em]
\includegraphics[width=0.65\columnwidth]{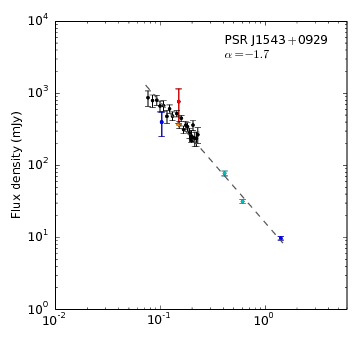}\hspace*{-0.5em}%                     
\hspace*{-0.5em}\includegraphics[width=0.65\columnwidth]{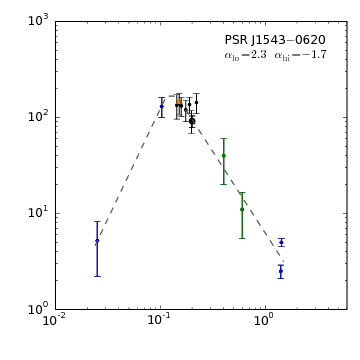}\hspace*{-0.5em}%     
\hspace*{-0.5em}\includegraphics[width=0.65\columnwidth]{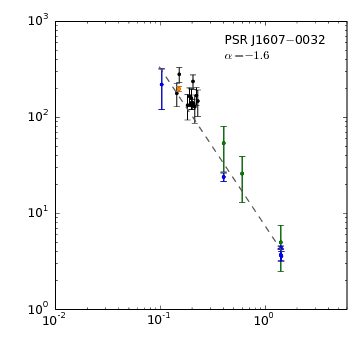} \\[-0.6em]
\includegraphics[width=0.65\columnwidth]{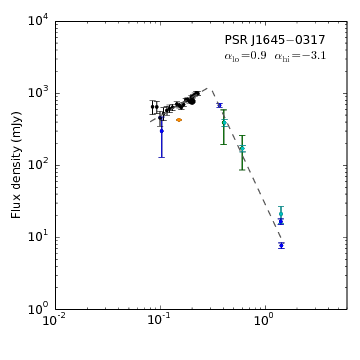}\hspace*{-0.5em}%                     
\hspace*{-0.5em}\includegraphics[width=0.65\columnwidth]{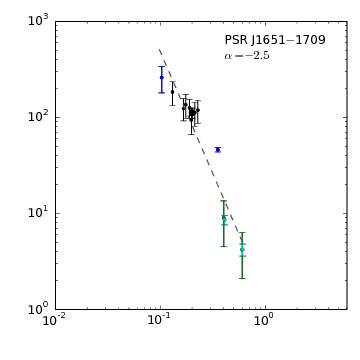}\hspace*{-0.5em}%     
\hspace*{-0.5em}\includegraphics[width=0.65\columnwidth]{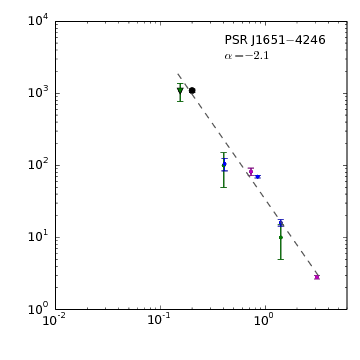} \\[-0.6em]
\includegraphics[width=0.65\columnwidth]{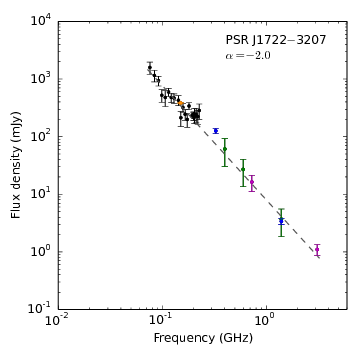}\hspace*{-0.5em}%                     
\hspace*{-0.5em}\includegraphics[width=0.65\columnwidth]{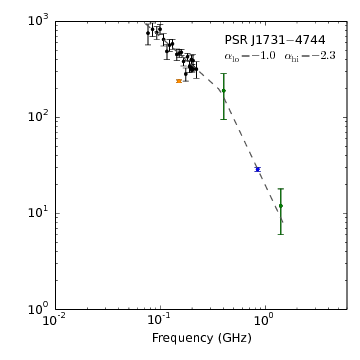}\hspace*{-0.5em}%     
\raisebox{0.9cm}{\hspace*{1.25cm}\includegraphics[width=0.4\columnwidth]{images/legend}}\\
\caption{(continued)}
\end{figure*}

\begin{figure*}\ContinuedFloat
\includegraphics[width=0.65\columnwidth]{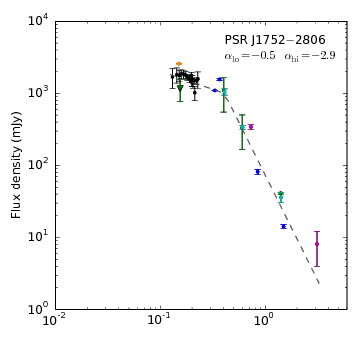}\hspace*{-0.5em}%                     
\hspace*{-0.5em}\includegraphics[width=0.65\columnwidth]{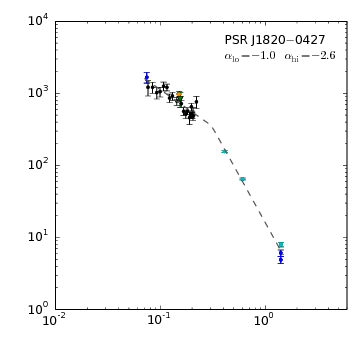}\hspace*{-0.5em}%     
\hspace*{-0.5em}\includegraphics[width=0.65\columnwidth]{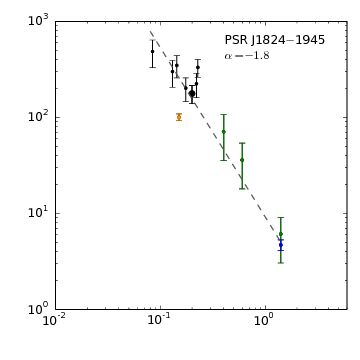} \\[-0.6em]
\includegraphics[width=0.65\columnwidth]{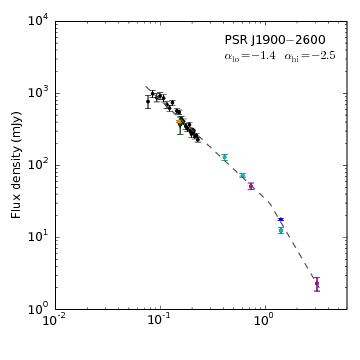}\hspace*{-0.5em}%                     
\hspace*{-0.5em}\includegraphics[width=0.65\columnwidth]{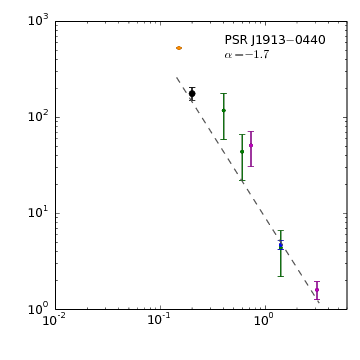}\hspace*{-0.5em}%     
\hspace*{-0.5em}\includegraphics[width=0.65\columnwidth]{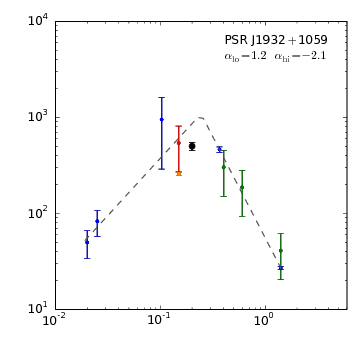} \\[-0.6em]
\includegraphics[width=0.65\columnwidth]{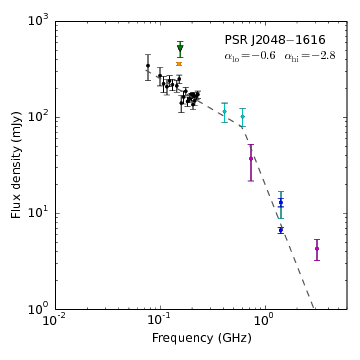}\hspace*{-0.5em}%                     
\hspace*{-0.5em}\includegraphics[width=0.65\columnwidth]{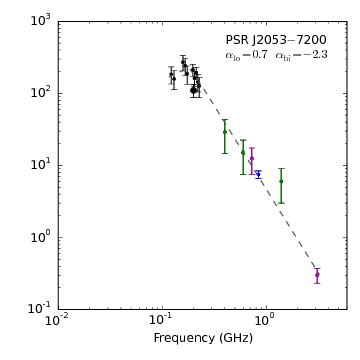}\hspace*{-0.5em}%     
\hspace*{-0.5em}\includegraphics[width=0.65\columnwidth]{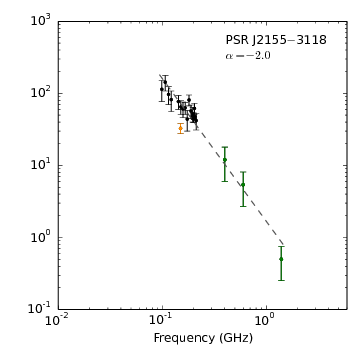} \\[-0.6em]
\raisebox{0.9cm}{\hspace*{1.25cm}\includegraphics[width=0.4\columnwidth]{images/legend}}\\
\caption{(continued)}
\end{figure*}

\begin{table*}
  \centering
  \caption{Flux density measurements and spectral indices for millisecond pulsars in our sample. The flux density at 200~MHz (S$_{200}$) is measured from mosaics that are averaged across the full 72--231~MHz bandwidth.
Sources identified as variable by \citet{bell16} are marked with (v). $N_{\rm p}$ is the number of measurements included in the fits.
The reference key is the same as for Table~\ref{sinormal}. Full SEDs are given in Table~4.}
  \label{similli}
  \begin{tabular}{llrccrr}
    \hline
    J name & B name & S$_{200}$  & Fit range & $N_{\rm p}$ & $\alpha_{\rm fit}$ & SED references \\
           &        & (mJy)        & (MHz)       & &     &   \\
    \hline
\input{milli_table}
    \hline
  \end{tabular}
\end{table*}

\begin{table*}
  \centering
  \caption{Fit results for sources where the spectrum was modelled by a broken power law.}
  \label{broken}
  \begin{tabular}{llccrr} % four columns, alignment for each                                         
    \hline
    J name & B name & Fit range & $\nu_{\rm br}$ & $\alpha_{\rm lo, fit}$ & $\alpha_{\rm hi, fit}$ \\
\
           &        & (MHz)       & (MHz) &    &     \\
    \hline
\input{pulsar_2pl}
    \hline
  \end{tabular}
\end{table*}

\subsection{Detected population}
\label{sec:detected}
\begin{figure}
\includegraphics[width=\columnwidth]{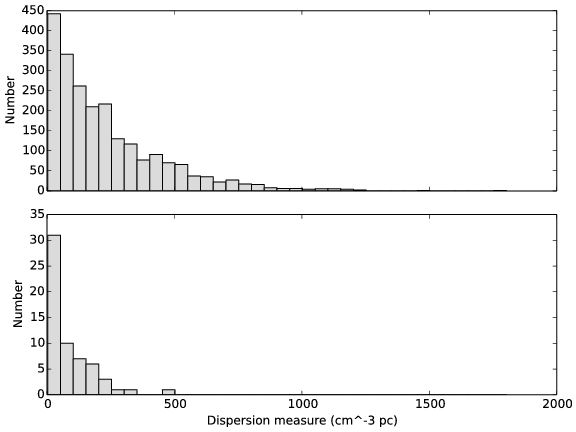}
\caption{Histogram of dispersion measures of all known pulsars in the GLEAM region (top panel) and the pulsars we detected (bottom panel).}\label{f_dm}
\end{figure}

Our survey will generally detect pulsars with high flux densities, which are likely to be closer, and hence have lower dispersion measures. This is demonstrated in Figure~\ref{f_dm}, which shows the pulsars we detect
typically have lower dispersion measures than the overall distribution.

\begin{figure}
\includegraphics[width=\columnwidth]{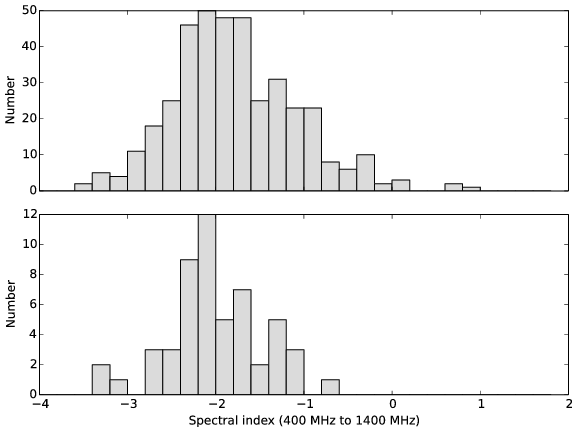}
\caption{Histogram of the spectral index $\alpha_{400}^{1400}$ measured between the ATNF catalogue listed values for
$S_{400}$ and $S_{1400}$ of all known pulsars in the GLEAM region (top panel) and for the pulsars we detected (bottom panel).}\label{f_si}
\end{figure}

There are 391 known pulsars that fall in the GLEAM survey region and have a flux density measurement at both 400~MHz (${\rm S}_{400}$) and 1.4~GHz (${\rm S}_{1400}$) listed in the ATNF pulsar catalogue. The distribution of spectral index, $\alpha_{400}^{1400}$, for these pulsars is shown in the top panel of Figure~\ref{f_si}. \changed{The mean spectral index of this distribution is $-1.8\pm0.7$. The distribution of spectral indices for
the subset of these that we detected is shown in the bottom panel of Figure~\ref{f_si}. This distribution has a mean spectral index of $-2.0\pm0.5$, which is steeper than the mean reported in the
literature of $\approx -1.6$ \citep{lorimer95}. This is expected given that our sample is selected based on detection at low frequency, which should prefer steep spectrum sources.}

\changed{To compare our detections rate with what would be expected at low frequencies we used the $S_{400}$ and $S_{1400}$ flux densities from the ATNF pulsar catalogue (where available) to predict the 200~MHz flux density.
In cases for which there was only a $S_{1400}$ flux density we assumed a spectral index of $-1.8$ (the median of the spectral index distribution). From these we were able to predict a 200~MHz flux density for 1559 pulsars in the GLEAM survey region. We considered 
a source detectable if it has a predicted flux density of $3\sigma = 45$~mJy~beam$^{-1}$ (based on the mean rms noise in the region of these pulsars of 15~mJy~beam$^{-1}$). 
With these limits we predicted that 61 sources would be detectable, which is extremely close to the number we detected: 60.
}

Using the ATNF pulsar catalogue flux densities and derived spectral indices we also looked at whether there were specific pulsars we would have expected to detect but did not. There are 38 sources with a predicted
$S_{200} > 45$~mJy~beam$^{-1}$ that were not in the sample of detected pulsars presented in this paper. Most of the non-detections near our limit were likely to be due to higher than average local rms noise, so we only considered sources above a $5\sigma$ threshold of 75~mJy~beam$^{-1}$, leaving 19 sources. We visually inspected the GLEAM maps at the positions of
all of these sources and found that either (i) the sources were detected, but had been excluded from our sample due because they were part of an extended structure or diffuse emission; or (ii) the sources had not
been detected and were in a region of higher than average noise, or in a negative bowl caused by imaging Galactic plane emission.

\subsection{Variability}\label{s_var}
Pulsars are known to exhibit variability, particularly at low frequencies when the effects of interstellar scintillation are stronger. In a related project \citep{bell16}, we identified four pulsars
that showed significant variability over timescales of minutes to months: PSR~J0034$-$0721, PSR~J0437$-$4715, PSR~J0630$-$2834 and PSR~J0953$+$0755. These are identified with a (v) in Tables~\ref{sinormal}--\ref{similli}. The last three of these four show high levels of scatter in their SEDs, in particular PSR~J0953$+$0755.

\begin{figure*}
\centering
\includegraphics[width=0.9\columnwidth]{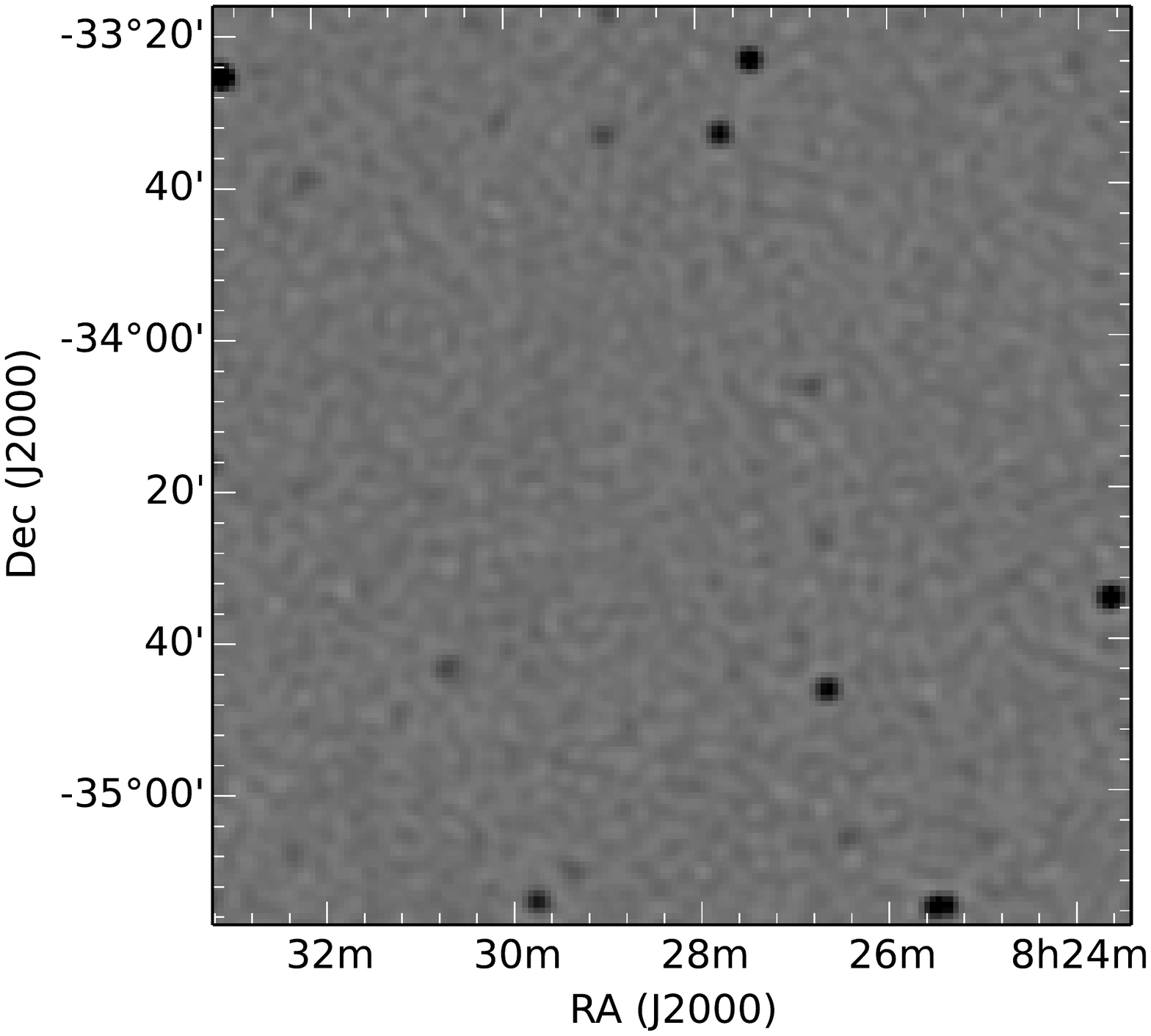}
\includegraphics[width=0.9\columnwidth]{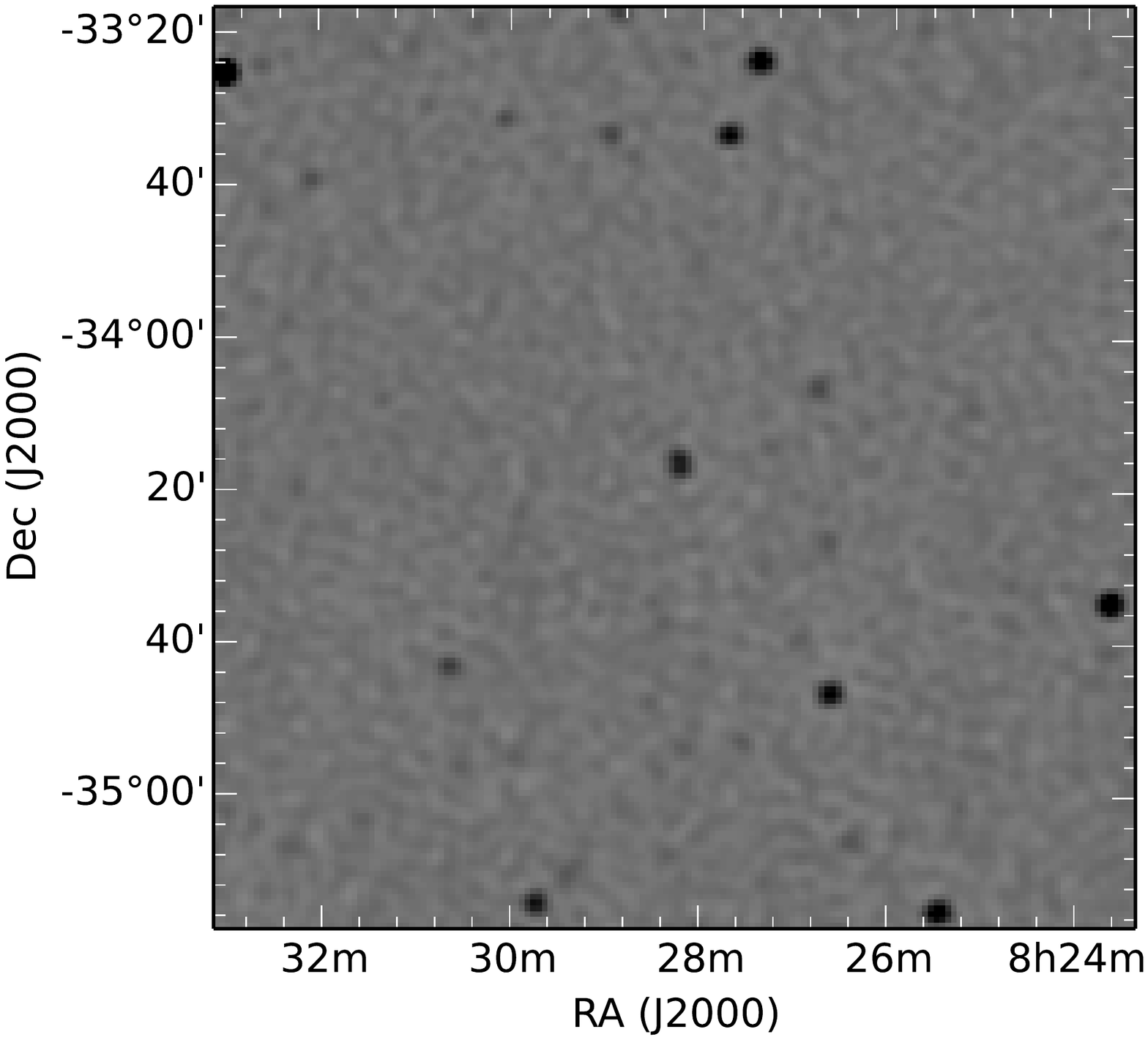} 
\vspace*{-1.5cm}
\caption{154~MHz images of PSR~J0828$-$3417 in its off (left) and on 
(right) states in two images from the MWA Transients Survey (MWATS; PI Bell). The two images are separated by 6 minutes: the image on the left was observed at 2016-02-01 15:53:36 UTC, and the image on the right was observed at 2016-02-01 15:59:36 UTC.}
\label{nulling}
\end{figure*}

Some pulsars also have significant intrinsic variability, which can also complicate broad-band SED measurement from non-contemporaneous images.
As an example, we show the intermittent pulsar PSR~J0828$-$3417, which has a reported duty cycle of $70\%$ \citep{durdin79,biggs85}. 
PSR~J0828$-$3417 switches between a strong mode and a weak mode with a typical timescales of hours \citep{esamdin12}. This pulsar was detected in the GLEAM images, although we only report a 200\,MHz flux density which is significantly above the other measurements.  To demonstrate that we have in fact measured the pulsar we show in Figure~\ref{nulling}  two individual 154~MHz MWA images (from the MWA Transients Survey; PI Bell) centred on the position of PSR~J0828$-$3417. In the first image (on the left) there is no detection of the source. In the second image, taken 6 minutes later, there is a clear detection with a measured flux density of 92~mJy~beam$^{-1}$.

\section{CONCLUSIONS}\label{s_conclusion}
We have presented new low-frequency flux density measurements for 60 pulsars from the ATNF pulsar catalogue. Our flux density measurements 
agree well with those previously reported in the literature and we find a median spectral index of $-1.8$ for the sources we detected.

Our analysis used data from the first year of the GLEAM survey, as discussed in \citet{wayth15} and \citet{hurleywalker16}.  We used images which were processed to optimize the high-latitude ($|b|>10\degree$) sky, so deconvolution of extended Galactic emission was not ideal.  Processing to probe more deeply into the Galactic plane is ongoing and should be released later this year, improving measurements of pulsars at low latitudes.  Processing of a second year of the GLEAM survey is also ongoing, which will allow searches for variability on timescales longer than those probed by \citet{bell16}.

Separately, our identification of pulsars was largely limited by the sensitivity of the survey, which is itself limited by confusion \citep{franzen16}.  The MWA has recently been upgraded to enable imaging with roughly double the current maximum baseline, up to $\approx 5\,$km.  This will result in a factor of 2 smaller FWHM of the point-spread function which will reduce the confusion level by a factor of $\sim 5$ \citep{franzen16}. Based on extrapolating the $S_{400}$ and $S_{1400}$ measurements from the ATNF pulsar catalogue, we predict that reducing the image confusion noise by a factor of 5 will increase the number of pulsars detected to approximately 200. Not only will this allow a better examination of the overall population, but it will allow more robust testing for correlations between the measured parameters and the intrinsic spin parameters of the pulsars.

\begin{acknowledgements}
We thank Fabian Jankowski for providing his ATCA flux density measurements ahead of publication.
This scientific work makes use of the Murchison Radio-astronomy Observatory, operated by CSIRO. We acknowledge the Wajarri Yamatji people as the traditional owners of the Observatory site. Support for the operation of the MWA is provided by the Australian Government Department of Industry and Science and Department of Education (National Collaborative Research Infrastructure Strategy: NCRIS), under a contract to Curtin University administered by Astronomy Australia Limited. We acknowledge the iVEC Petabyte Data Store and the Initiative in Innovative Computing and the CUDA Center for Excellence sponsored by NVIDIA at Harvard University.  DLK and SDC are additionally supported by NSF grant AST-1412421. 

This research has made use of the NASA/IPAC Extragalactic Database (NED) which is operated by the Jet Propulsion Laboratory, California Institute of Technology, under contract with the National Aeronautics and Space Administration. It has also made use of the SIMBAD database, operated at CDS, Strasbourg, France.
\end{acknowledgements}

\bibliographystyle{apj}
\bibliography{mn-jour,mwa,pulsars,vast}

\include{maintable}

\end{document}

%% file: pulsar_1pl.tex
J0034$-$0721 (v) & B0031$-$07 & $292 \pm 14$ & 20--1400 & 26 & 2pl & Be16,C98,F16,M00,T93,Z13 \\
J0206$-$4028 & B0203$-$40 & $32 \pm 6$ & 107--1400 & 8 & $-1.7\pm0.1$ & M78,T93 \\
J0452$-$1759 & B0450$-$18 & $96 \pm 7$ & 130--1408 & 16 & 2pl & D02,F16,L95,T93 \\
J0630$-$2834 (v) & B0628$-$28 & $463 \pm 5$ & 74--3100 & 11 & 2pl & Be16,C07,C98,D96,F16,J16,L95,T93 \\
J0738$-$4042 & B0736$-$40 & $165 \pm 13$ & 107--8400 & 20 & 2pl & F16,J05,J15,M07,M78 \\
J0809$-$4753 & B0808$-$47 & $229 \pm 14$ & 200--3100 & 5 & $-2.2\pm0.1$ & F16,J16,T93 \\
J0820$-$4114 & B0818$-$41 & $116 \pm 16$ & 76--3100 & 23 & 2pl & F16,J16,M07,T93 \\
J0820$-$1350 & B0818$-$13 & $160 \pm 7$ & 76--3100 & 27 & $-1.6\pm0.1$ & C98,F16,J16,L95 \\
J0826$+$2637 & B0823$+$26 & $243 \pm 21$ & 20--14800 & 23 & 2pl & B16,B78,C98,F16,M00,T93,Z13 \\
J0828$-$3417 (v) & B0826$-$34 & $400 \pm 8$ & -- & -- & -- & H04,I16,T93 \\
J0835$-$4510 & B0833$-$45 & $7075 \pm 207$ & 80--3100 & 9 & 2pl & Be16,F16,H14,J16,S95,T93 \\
J0837$+$0610 & B0834$+$06 & $286 \pm 13$ & 20--1400 & 10 & 2pl & C98,F16,J16,L14,M00,T93,Z13 \\
J0837$-$4135 & B0835$-$41 & $95 \pm 16$ & 115--3100 & 21 & 2pl & F16,J16,M07,T93 \\
J0840$-$5332 & B0839$-$53 & $56 \pm 13$ & 115--1400 & 16 & $-1.6\pm0.1$ & J16,T93 \\
J0855$-$3331 & B0853$-$33 & $47 \pm 8$ & 130--1408 & 14 & $-1.9\pm0.1$ & F16,J16,L95,T93 \\
J0856$-$6137 & B0855$-$61 & $85 \pm 9$ & 200--1382 & 4 & $-1.6\pm0.1$ & J16,T93 \\
J0905$-$5127 & $-$ & $73 \pm 15$ & 92--1400 & 18 & $-2.0\pm0.1$ & H04,I16,T93 \\
J0907$-$5157 & B0905$-$51 & $106 \pm 11$ & 84--3100 & 26 & 2pl & Bh16,F16,J16,M07,T93 \\
J0922$+$0638 & B0919$+$06 & $100 \pm 13$ & 20--3100 & 27 & 2pl & C98,F16,J16,T93,Z13 \\
J0924$-$5302 & B0922$-$52 & $96 \pm 9$ & 99--3100 & 24 & $-2.0\pm0.1$ & Bh16,J16,M07,T93 \\
J0942$-$5552 & B0940$-$55 & $73 \pm 14$ & 151--3100 & 15 & 2pl & J16,M07,T93 \\
J0942$-$5657 & B0941$-$56 & $112 \pm 11$ & 400--3100 & 6 & $-1.5\pm1.9$ & J16,M07,M78,T93 \\
J0953$+$0755 (v) & B0950$+$08 & $1072 \pm 17$ & 20--3100 & 36 & 2pl & Be16,C07,C98,D96,F16 \\
             &                &               &          &    &     & J16,L95,M00,S95,T93,Z13 \\
J0959$-$4809 & B0957$-$47 & $50 \pm 7$ & 151--3100 & 9 & 2pl & F16,J16,T93 \\
J1001$-$5507 & B0959$-$54 & $142 \pm 12$ & 400--3100 & 6 & $-1.7\pm0.1$ & J16,M07,T93 \\
J1012$-$2337 & B1010$-$23 & $47 \pm 40$ & 200--600 & 4 & $-3.2\pm0.2$ & L95,T93 \\
J1047$-$3032 & $-$ & $24 \pm 5$ & 111--1400 & 5 & $-5.9\pm0.1$ & G98,L98 \\
J1057$-$5226 & B1055$-$52 & $202 \pm 12$ & 150--843 & 4 & $-1.5\pm0.1$ & Be16,I16,M07,T93 \\
J1116$-$4122 & B1114$-$41 & $52 \pm 7$ & 92--3100 & 19 & 2pl & F16,J16,T93 \\
J1121$-$5444 & B1119$-$54 & $101 \pm 15$ & 122--1400 & 15 & $-2.1\pm0.1$ & H04,J16,T93 \\
J1136$+$1551 & B1133$+$16 & $684 \pm 61$ & 20--1408 & 33 & 2pl & B16,C07,C98,D96,F16 \\
             &                &               &          &    &     & J16,L95,M00,T93,Z13 \\
J1327$-$6222 & B1323$-$62 & $284 \pm 48$ & 143--3100 & 17 & 2pl & J16,M07,T93 \\
J1359$-$6038 & B1356$-$60 & $402 \pm 94$ & 115--1400 & 21 & 2pl & Be16,M07,M78,N08,T93 \\
J1430$-$6623 & B1426$-$66 & $190 \pm 28$ & 130--3100 & 18 & 2pl & H04,J15,M07,T93 \\
J1453$-$6413 & B1449$-$64 & $684 \pm 23$ & 76--1400 & 24 & 2pl & Be16,M07,N08,T93 \\
J1456$-$6843 & B1451$-$68 & $738 \pm 21$ & 76--1400 & 24 & 2pl & Be16,M07,T93 \\

%% file: pulsar_1plb.tex
J1543$-$0620 & B1540$-$06 & $91 \pm 12$ & 25--1420 & 13 & 2pl & C98,F16,M00,Se95,T93,Z13 \\
J1543$+$0929 & B1541$+$09 & $234 \pm 19$ & 76--1400 & 25 & $-1.7\pm0.1$ & B16,C98,F16,L95,M00 \\
J1607$-$0032 & B1604$-$00 & $137 \pm 15$ & 102--1420 & 17 & $-1.6\pm0.1$ & B95,C98,F16,M00,M78,Se95,T93 \\
J1645$-$0317 & B1642$-$03 & $774 \pm 18$ & 84--1420 & 28 & 2pl & C98,D96,F16,L95,M00,Se95,T93 \\
J1651$-$1709 & B1648$-$17 & $111 \pm 13$ & 102--606 & 13 & $-2.5\pm0.3$ & D02,L95,M00,T93 \\
J1651$-$4246 & B1648$-$42 & $1095 \pm 53$ & 154--3100 & 8 & $-2.1\pm0.2$ & Be16,H04,J16,M07,M78,T93 \\
J1722$-$3207 & B1718$-$32 & $229 \pm 37$ & 76--3100 & 25 & $-2.0\pm0.1$ & F16,H04,J16,R97,T93 \\
J1731$-$4744 & B1727$-$47 & $325 \pm 28$ & 76--1400 & 21 & 2pl & F16,M07,T93 \\
J1752$-$2806 & B1749$-$28 & $1504 \pm 269$ & 130--3100 & 26 & 2pl & Be16,D96,F16,H92,J16 \\
             &                &               &          &    &     & L95,M07,N04,T93\\
J1820$-$0427 & B1818$-$04 & $499 \pm 51$ & 74--1408 & 25 & 2pl & Be16,C98,F16,H04,L14,L95 \\
J1824$-$1945 & B1821$-$19 & $177 \pm 38$ & 84--1400 & 10 & $-1.8\pm0.1$ & C98,F16,T93 \\
J1900$-$2600 & B1857$-$26 & $299 \pm 13$ & 76--3100 & 27 & 2pl & Be16,C98,F16,J16,L95 \\
J1913$-$0440 & B1911$-$04 & $176 \pm 26$ & 400--3100 & 6 & $-1.7\pm0.4$ & C98,F16,J16,T93 \\
J1932$+$1059 & B1929$+$10 & $501 \pm 47$ & 20--1400 & 9 & 2pl & B16,C98,D96,F16,M00,T93,Z13 \\
J2048$-$1616 & B2045$-$16 & $169 \pm 8$ & 76--3100 & 26 & 2pl & Be16,C98,F16,J16,L95,N08 \\
J2053$-$7200 & B2048$-$72 & $110 \pm 22$ & 122--3100 & 16 & 2pl & J16,M03,T93 \\
J2155$-$3118 & B2152$-$31 & $46 \pm 6$ & 99--1400 & 17 & $-2.0\pm0.1$ & F16,T93 \\

%% file: milli_table.tex
J0034$-$0534 & $-$ & $65 \pm 11$ & 74--1660 & 25 & $-2.6\pm0.1$ & F16,K15,K16,L14,T93,T98 \\
J0437$-$4715 (v) & $-$ & $834 \pm 9$ & 76--3100 & 33 & -- & Be16,D15,F16,J16,J93,M03,M96 \\
J0737$-$3039A & $-$ & $53 \pm 8$ & 99--1400 & 13 & $-1.4\pm0.1$ & C98,F16,K16 \\
J1643$-$1224 & $-$ & $123 \pm 14$ & 107--1400 & 17 & $-1.6\pm0.1$ & C98,D02,F16,L95,M13 \\
J1810$+$1744 & $-$ & $231 \pm 35$ & 74--350 & 15 & $-2.3\pm0.2$ & F16,H11,K15,K16,L14 \\
J1824$-$2452A & B1821$-$24A & $199 \pm 27$ & 76--1400 & 21 & $-3.2\pm0.1$ & F16,H04,T93 \\
J2241$-$5236 & $-$ & $60 \pm 11$ & 115--1400 & 13 & $-1.3\pm0.1$ & K11,M03 \\

%% file: pulsar_2pl.tex
J0034$-$0721 (v) & B0031$-$07 & 20--1400 & $77\pm5$ & $0.8\pm0.3$ & $-1.4\pm0.1$ \\
J0452$-$1759 & B0450$-$18 & 130--1408 & $606\pm25$ & $-0.3\pm0.1$ & $-2.9\pm0.2$ \\
J0630$-$2834 (v) & B0628$-$28 & 74--3100 & $364\pm252$ & $-1.2\pm0.6$ & $-2.3\pm0.1$ \\
J0738$-$4042 & B0736$-$40 & 107--8400 & $772\pm63$ & $0.2\pm0.1$ & $-1.4\pm0.1$ \\
J0820$-$4114 & B0818$-$41 & 76--3100 & $444\pm86$ & $-0.7\pm0.2$ & $-2.0\pm0.2$ \\
J0826$+$2637 & B0823$+$26 & 20--14800 & $263\pm17$ & $0.2\pm0.1$ & $-2.0\pm0.1$ \\
J0835$-$4510 & B0833$-$45 & 80--3100 & $664\pm141$ & $-0.5\pm0.2$ & $-1.9\pm0.1$ \\
J0837$+$0610 & B0834$+$06 & 20--1400 & $143\pm10$ & $0.8\pm0.1$ & $-2.6\pm0.1$ \\
J0837$-$4135 & B0835$-$41 & 115--3100 & $987\pm246$ & $-0.3\pm0.1$ & $-1.8\pm0.5$ \\
J0907$-$5157 & B0905$-$51 & 84--3100 & $431\pm93$ & $-0.7\pm0.1$ & $-1.2\pm0.1$ \\
J0922$+$0638 & B0919$+$06 & 20--3100 & $131\pm12$ & $0.1\pm0.3$ & $-1.9\pm0.1$ \\
J0942$-$5552 & B0940$-$55 & 151--3100 & $533\pm140$ & $-0.8\pm0.2$ & $-1.7\pm0.1$ \\
J0953$+$0755 (v) & B0950$+$08 & 20--3100 & $109\pm8$ & $1.9\pm1.4$ & $-1.5\pm0.2$ \\
J0959$-$4809 & B0957$-$47 & 151--3100 & $1382\pm202$ & $-1.1\pm0.1$ & $-3.0\pm0.2$ \\
J1116$-$4122 & B1114$-$41 & 92--3100 & $1381\pm308$ & $-1.3\pm0.2$ & $-3.2\pm0.4$ \\
J1136$+$1551 & B1133$+$16 & 20--1408 & $210\pm5$ & $0.6\pm0.1$ & $-2.1\pm0.1$ \\
J1327$-$6222 & B1323$-$62 & 143--3100 & $322\pm52$ & $0.2\pm0.5$ & $-1.9\pm0.2$ \\
J1359$-$6038 & B1356$-$60 & 115--1400 & $220\pm25$ & $-0.3\pm0.4$ & $-2.1\pm0.1$ \\
J1430$-$6623 & B1426$-$66 & 130--3100 & $605\pm243$ & $-0.5\pm0.7$ & $-2.8\pm0.3$ \\
J1453$-$6413 & B1449$-$64 & 76--1400 & $363\pm35$ & $-0.3\pm0.1$ & $-2.8\pm0.2$ \\
J1456$-$6843 & B1451$-$68 & 76--1400 & $234\pm140$ & $0.0\pm0.1$ & $-2.2\pm1.0$ \\
J1543$-$0620 & B1540$-$06 & 25--1420 & $125\pm15$ & $2.3\pm0.7$ & $-1.7\pm0.1$ \\
J1645$-$0317 & B1642$-$03 & 84--1420 & $297\pm8$ & $0.9\pm0.1$ & $-3.1\pm0.1$ \\
J1731$-$4744 & B1727$-$47 & 76--1400 & $372\pm162$ & $-1.0\pm0.1$ & $-2.3\pm0.7$ \\
J1752$-$2806 & B1749$-$28 & 130--3100 & $407\pm58$ & $-0.5\pm0.3$ & $-2.9\pm0.3$ \\
J1820$-$0427 & B1818$-$04 & 74--1408 & $306\pm32$ & $-1.0\pm0.2$ & $-2.6\pm0.1$ \\
J1900$-$2600 & B1857$-$26 & 76--3100 & $1091\pm162$ & $-1.4\pm0.1$ & $-2.5\pm0.5$ \\
J1932$+$1059 & B1929$+$10 & 20--1400 & $243\pm25$ & $1.2\pm0.2$ & $-2.1\pm0.1$ \\
J2048$-$1616 & B2045$-$16 & 76--3100 & $605\pm128$ & $-0.6\pm0.2$ & $-2.8\pm0.5$ \\
J2053$-$7200 & B2048$-$72 & 122--3100 & $185\pm10$ & $0.7\pm0.6$ & $-2.3\pm0.1$ \\

%% file: maintable.tex
\begin{onecolumn}
\begin{sidewaystable}
\vspace*{2cm}
{\bf Table 4.} MWA flux density measurements or $3\sigma$ limits for the first 15 sources in our sample. The full table is available online, and in the online version
limits are marked with an L in the error columns.\\

\setlength{\tabcolsep}{1mm}
  \label{t_seds}
    \begin{tabular}{llrrrrrrrrrrr}
    \hline
    \input{pulsar_sed1}    
    \hline
    & & & & & & & & & & & & \\ 
    & & & & & & & & & & & & \\ 
    & & & & & & & & & & & & \\ 
    \hline
    \input{pulsar_sed2}
    \hline
    \end{tabular}
\end{sidewaystable}
\end{onecolumn}

%% file: pulsar_sed1.tex
J name & B name & \thead{S$_{200}$} & \thead{S$_{76}$} & \thead{S$_{84}$} & \thead{S$_{92}$} & \thead{S$_{99}$} & \thead{S$_{107}$} & \thead{S$_{115}$} & \thead{S$_{122}$} & \thead{S$_{130}$} & \thead{S$_{143}$}\\
 &  & \thead{(mJy)} & \thead{(mJy)} & \thead{(mJy)} & \thead{(mJy)} & \thead{(mJy)} & \thead{(mJy)} & \thead{(mJy)} & \thead{(mJy)} & \thead{(mJy)} & \thead{(mJy)}\\ \hline
J0034$-$0721 (v) & B0031$-$07 & $292 \pm 14$ & $923 \pm 95$ & $765 \pm 74$ & $693 \pm 64$ & $556 \pm 59$ & $694 \pm 46$ & $524 \pm 36$ & $447 \pm 35$ & $411 \pm 33$ & $424 \pm 26$ \\
J0034$-$0534 & $-$ & $65 \pm 11$ & $1406 \pm 85$ & $1032 \pm 69$ & $742 \pm 61$ & $762 \pm 58$ & $545 \pm 48$ & $420 \pm 40$ & $410 \pm 40$ & $303 \pm 33$ & $232 \pm 26$ \\
J0206$-$4028 & B0203$-$40 & $32 \pm 6$ & $<171$ & $<123$ & $<113$ & $<99$ & $86 \pm 27$ & $<72$ & $<60$ & $<51$ & $<51$ \\
J0437$-$4715 (v) & $-$ & $834 \pm 9$ & $1844 \pm 79$ & $1638 \pm 59$ & $1680 \pm 55$ & $1488 \pm 50$ & $1355 \pm 41$ & $1205 \pm 34$ & $1093 \pm 34$ & $953 \pm 28$ & $940 \pm 21$ \\
J0452$-$1759 & B0450$-$18 & $96 \pm 7$ & $<237$ & $<168$ & $<159$ & $<180$ & $<113$ & $<105$ & $<87$ & $156 \pm 28$ & $<78$ \\
J0630$-$2834 (v) & B0628$-$28 & $463 \pm 5$ & $824 \pm 81$ & $855 \pm 59$ & $1101 \pm 55$ & $1016 \pm 57$ & $1078 \pm 35$ & $1030 \pm 27$ & $992 \pm 25$ & $937 \pm 22$ & $874 \pm 22$ \\
J0737$-$3039A & $-$ & $53 \pm 8$ & $<264$ & $<198$ & $<195$ & $215 \pm 66$ & $237 \pm 45$ & $247 \pm 38$ & $<102$ & $<96$ & $90 \pm 23$ \\
J0738$-$4042 & B0736$-$40 & $165 \pm 13$ & $<261$ & $<204$ & $<186$ & $<195$ & $168 \pm 48$ & $<126$ & $<107$ & $144 \pm 38$ & $128 \pm 28$ \\
J0809$-$4753 & B0808$-$47 & $229 \pm 14$ & $370 \pm 74$ & $227 \pm 57$ & $190 \pm 50$ & $215 \pm 47$ & $300 \pm 43$ & $305 \pm 33$ & $262 \pm 30$ & $315 \pm 28$ & $249 \pm 21$ \\
J0820$-$4114 & B0818$-$41 & $116 \pm 16$ & $212 \pm 68$ & $<165$ & $<132$ & $<141$ & $189 \pm 38$ & $175 \pm 37$ & $<93$ & $148 \pm 33$ & $137 \pm 29$ \\
J0820$-$1350 & B0818$-$13 & $160 \pm 7$ & $751 \pm 69$ & $605 \pm 54$ & $391 \pm 44$ & $357 \pm 42$ & $454 \pm 36$ & $312 \pm 33$ & $288 \pm 28$ & $266 \pm 25$ & $282 \pm 23$ \\
J0826$+$2637 & B0823$+$26 & $243 \pm 21$ & $612 \pm 161$ & $<387$ & $<354$ & $<345$ & $<528$ & $487 \pm 147$ & $538 \pm 122$ & $<396$ & $403 \pm 57$ \\
J0828$-$3417 (v) & B0826$-$34 & $400 \pm 8$ & $198 \pm 64$ & $256 \pm 50$ & $154 \pm 39$ & $251 \pm 37$ & $325 \pm 35$ & $325 \pm 31$ & $373 \pm 28$ & $394 \pm 26$ & $434 \pm 18$ \\
J0835$-$4510 & B0833$-$45 & $7075 \pm 207$ & $10498 \pm 723$ & $9505 \pm 676$ & $8949 \pm 596$ & $8371 \pm 531$ & $8048 \pm 440$ & $7762 \pm 377$ & $7578 \pm 325$ & $7511 \pm 262$ & $7415 \pm 208$ \\
J0837$+$0610 & B0834$+$06 & $286 \pm 13$ & $588 \pm 91$ & $611 \pm 74$ & $574 \pm 63$ & $516 \pm 58$ & $515 \pm 62$ & $517 \pm 52$ & $533 \pm 47$ & $518 \pm 48$ & $499 \pm 33$ \\

%% file: pulsar_sed2.tex
J name & B name & \thead{S$_{151}$} & \thead{S$_{158}$} & \thead{S$_{166}$} & \thead{S$_{174}$} & \thead{S$_{181}$} & \thead{S$_{189}$} & \thead{S$_{197}$} & \thead{S$_{151}$} & \thead{S$_{212}$} & \thead{S$_{220}$} & \thead{S$_{227}$}\\
 &  & \thead{(mJy)} & \thead{(mJy)} & \thead{(mJy)} & \thead{(mJy)} & \thead{(mJy)} & \thead{(mJy)} & \thead{(mJy)} & \thead{(mJy)} & \thead{(mJy)} & \thead{(mJy)} & \thead{(mJy)}\\ \hline
J0034$-$0721 (v) & B0031$-$07 & $368 \pm 22$ & $337 \pm 21$ & $316 \pm 21$ & $299 \pm 21$ & $252 \pm 19$ & $309 \pm 19$ & $263 \pm 19$ & $368 \pm 22$ & $277 \pm 20$ & $277 \pm 20$ & $254 \pm 21$ \\
J0034$-$0534 & $-$ & $174 \pm 22$ & $130 \pm 23$ & $76 \pm 21$ & $<66$ & $106 \pm 20$ & $63 \pm 18$ & $114 \pm 19$ & $174 \pm 22$ & $<66$ & $64 \pm 21$ & $<66$ \\
J0206$-$4028 & B0203$-$40 & $<51$ & $<48$ & $<45$ & $66 \pm 16$ & $<39$ & $<39$ & $<39$ & $<51$ & $32 \pm 9$ & $33 \pm 9$ & $<26$ \\
J0437$-$4715 (v) & $-$ & $947 \pm 19$ & $950 \pm 17$ & $861 \pm 15$ & $843 \pm 19$ & $731 \pm 16$ & $1329 \pm 16$ & $1307 \pm 17$ & $947 \pm 19$ & $490 \pm 19$ & $525 \pm 19$ & $501 \pm 20$ \\
J0452$-$1759 & B0450$-$18 & $88 \pm 21$ & $96 \pm 19$ & $<53$ & $111 \pm 16$ & $78 \pm 14$ & $86 \pm 14$ & $65 \pm 14$ & $88 \pm 21$ & $98 \pm 14$ & $100 \pm 13$ & $106 \pm 14$ \\
J0630$-$2834 (v) & B0628$-$28 & $772 \pm 19$ & $725 \pm 20$ & $714 \pm 18$ & $689 \pm 15$ & $610 \pm 13$ & $582 \pm 12$ & $484 \pm 11$ & $772 \pm 19$ & $419 \pm 11$ & $375 \pm 10$ & $368 \pm 10$ \\
J0737$-$3039A & $-$ & $<72$ & $<63$ & $96 \pm 20$ & $72 \pm 15$ & $68 \pm 14$ & $58 \pm 14$ & $49 \pm 13$ & $<72$ & $<42$ & $65 \pm 12$ & $67 \pm 12$ \\
J0738$-$4042 & B0736$-$40 & $123 \pm 25$ & $174 \pm 24$ & $133 \pm 22$ & $103 \pm 22$ & $205 \pm 19$ & $149 \pm 17$ & $136 \pm 18$ & $123 \pm 25$ & $148 \pm 13$ & $156 \pm 12$ & $172 \pm 11$ \\
J0809$-$4753 & B0808$-$47 & $267 \pm 20$ & $263 \pm 18$ & $263 \pm 15$ & $265 \pm 17$ & $260 \pm 18$ & $229 \pm 15$ & $217 \pm 14$ & $267 \pm 20$ & $207 \pm 16$ & $169 \pm 15$ & $171 \pm 18$ \\
J0820$-$4114 & B0818$-$41 & $133 \pm 23$ & $141 \pm 20$ & $116 \pm 22$ & $88 \pm 25$ & $158 \pm 23$ & $112 \pm 21$ & $138 \pm 19$ & $133 \pm 23$ & $125 \pm 16$ & $84 \pm 12$ & $113 \pm 12$ \\
J0820$-$1350 & B0818$-$13 & $220 \pm 22$ & $179 \pm 20$ & $182 \pm 20$ & $188 \pm 22$ & $237 \pm 20$ & $151 \pm 18$ & $143 \pm 18$ & $220 \pm 22$ & $175 \pm 14$ & $108 \pm 17$ & $178 \pm 15$ \\
J0826$+$2637 & B0823$+$26 & $375 \pm 49$ & $345 \pm 50$ & $417 \pm 52$ & $490 \pm 58$ & $336 \pm 56$ & $475 \pm 51$ & $293 \pm 51$ & $375 \pm 49$ & $397 \pm 68$ & $384 \pm 69$ & $448 \pm 94$ \\
J0828$-$3417 (v) & B0826$-$34 & $455 \pm 17$ & $406 \pm 15$ & $430 \pm 14$ & $448 \pm 16$ & $401 \pm 15$ & $397 \pm 13$ & $314 \pm 12$ & $455 \pm 17$ & $292 \pm 10$ & $285 \pm 10$ & $258 \pm 10$ \\
J0835$-$4510 & B0833$-$45 & $7274 \pm 197$ & $7194 \pm 178$ & $7131 \pm 162$ & $7075 \pm 138$ & $6871 \pm 115$ & $6755 \pm 99$ & $6578 \pm 109$ & $7274 \pm 197$ & $6259 \pm 85$ & $6228 \pm 99$ & $5790 \pm 79$ \\
J0837$+$0610 & B0834$+$06 & $433 \pm 27$ & $364 \pm 26$ & $388 \pm 26$ & $354 \pm 27$ & $349 \pm 26$ & $370 \pm 27$ & $359 \pm 27$ & $433 \pm 27$ & $326 \pm 30$ & $282 \pm 31$ & $284 \pm 33$ \\